# Medium range real atomic structure of face centred icosahedral $Ho_9Mg_{26}Zn_{65}$

S. Brühne, R. Sterzel, E. Uhrig, C. Gross, and W. Assmus

*Physikalisches Institut, Johann Wolfgang Goethe-Universität, Robert-Mayer-Str. 2-4
D-60054 Frankfurt/Main, Germany*
(to appear in Phys. Rev. *B*, 2003)

A new approach to solve quasicrystalline atomic structures in 3-dimensional (3D) real space is presented: The atomic pair distribution function (PDF) of face centred icosahedral $Ho_9Mg_{26}Zn_{65}$ [$a(6D) = 2 \times 5.18(3)$Å] was obtained from in-house X-ray diffraction data (MoK$\alpha_1$). Starting with rational approximant models, derived from 1/1- and 2/1-Al-Mg-Zn, its local and medium range structure was refined ($r < 27$Å; $R = 12.9\%$) using the PDF data. 85% of all atoms show Frank-Kasper (FK) type coordinations. Basic structural unit is the 3-shell, 104-atom Bergman cluster ($d \approx 15$Å) comprising a void at its center. The clusters are interconnected sharing common edges and hexagonal faces of the 3rd shells. The remaining space is filled by some glue atoms (9% of all atoms), yielding an almost tetrahedrally close packed structure. All Ho atoms are surrounded by 16 neighbours (FK-polyhedron "P"). Most of them (89%) are situated in the 2nd shell (pentagon dodecahedron), the other act as glue atoms. As a result and as can be expected for real matter, local atomic coordinations in quasicrystals are similar compared to common crystalline intermetallic compounds. From our results, the long range quasiperiodic structure of icosahedral Mg-Zn-*RE* (*RE* = Y and some rare earths) is anticipated to be a canonical cell tiling (CCT, after Henely) decorated with Bergman clusters.

## I. INTRODUCTION

Aperiodic crystals with icosahedral diffraction symmetry — icosahedral (or short: *i*) quasicrystals — in the Mg-Zn-*RE* systems (*RE* = Y, Gd, Tb, Dy, Ho, Er) were discovered in 1993.[1] Since they contain 4*f* elements, they are potential candidates for quasiperiodic long range magnetic order. Basing on the determination of the primary cristallisation field of face centred icosahedral (*fci*)-Mg-Y-Zn,[2] large and well ordered single crystals of *fci*-Mg-Zn-*RE* (*RE* = Dy, Er, Ho, Tb, Y)[3,4] are available. So a variety of physical properties, such as ultrasonic behaviour[5], diffusion[6], optical[7] or electronic features[8,9], have been investigated. But still the atomic structure of *fci*-Zn-Mg-*RE* is not clear. However, as macroscopic properties are governed by the microscopic structure, its knowledge is indispensable in order to gain insight in the structure property relations of these materials. Beyond, structural information is the basis to be able to taylor materials in order to tune their properties.

Today, crystal structure determination of periodic crystals is a more or less standard technique using well established and innumerably proven methods. It firmly bases on periodicity as a hard constraint. However, structure determination of quasiperiodic crystals (quasicrystals), even almost 20 years after their discovery in 1984 by Shechtman *et al.*[10], is still no straightforward task. A number of models comprising various tilings and their decorations were developed but they still lack of thorough experimental evidence comparable to that of periodic crystals. Most approaches to solve and refine quasiperiodic structures are based on higher dimensional crystallography (*n*-dimensional: $n$D; $n > 3$) using single crystal diffraction data (5D, decagonal or 6D, icosahedral phases)[11]. In the icosahedral case, a 6D periodic hypercubic crystal yields a 3D quasiperiodic real space atomic structure *via* a sophisticated cut-and-projection method. Though a definite set of parameters as a 6D lattice constant $a(6D) = 1/a^*_{100000}$, 6D superspace group and 6D coordinates of hyperatoms and their 3D surfaces in a perpendicular subspace describe well the periodocity in six dimensions — but it is not trivial to derive its 3D parallel (or physical) real space meaning. So still the felicitous statement of Shoemaker and Shoemaker of 1988 holds: "[…] crystallographers and chemists, […], they will not be satisfied until detailed atomic structure models – with real atoms, credibly coordinated, at credible interatomic distances – have been solidly established from obtained diffraction intensities."[12]

Our effort presented here is a new route to circumvent the problems posed by the absence of periodicity. The analysis of the 1D atomic pair distribution function (PDF), easily accessible from powder diffraction data, does not rely on information on whatever periodicity. The PDF reflects the *local* structure, *i.e.* the probability to find any atom at a given distance to any other atom of the diffracting sample, averaged over all atoms; see equation 2. In the early days of quasicrystal investigation already, qualitative short range similarities to related known crystalline phases have been shown by calculating the PDF[13]. Traditionally, pair distribution functions are used for investigation of glasses and amorphous materials[14] or, more recently, to quantify disorder in periodic crystals from diffuse scattering[15]. To our knowledge, this paper describes for first time a real space least squares refinement of a PDF obtained from powder diffraction data of quasicrystalline material. To perform such a refinement, an unevitable prerequisite is a good 3D atomic structure model as a starting point. Both finding the starting model by "crystal chemical intuition" and its verification by refinement will be subject of this work.

The paper at hand is structured as followed: In section II known (quasi)crystal structures in the Mg-Zn-*RE* systems are discussed to get an idea about suitable starting models. With it, the concept of the rational approximant, a classification of *i* quasicrystals and the Frank Kasper concept are introduced. Then, an experimental and methods section (III) follows on the synthesis of *fci*-Ho-Mg-Zn and how to obtain the PDF from powder diffraction. In section IV, this experimental PDF is compared as a fingerprint to simulated PDFs of crystalline phases of Mg-Zn-*RE* and other systems. Finally it turns out that starting models based on a sequence of rational approximants in the Mg-Zn-*B* (*B* = Al,Ga) systems can be constructed. These models are refined in section V. The refined data reflect the short and medium range atomic arrangement in $fci$-$Ho_9Mg_{26}Zn_{65}$, the data are discussed in section VI in detail. Section VII then concludes the paper.

## II. THE Mg-Zn-*RE* SYSTEMS

*Quasicrystalline phases*

In the ternary systems Mg-Zn-*RE* (*RE* = Dy, Er, Gd, Ho, Lu, Nd, Sm, Tm and Y) a number of quasicrystalline phases is known: (**1**) face centred icosahedral (*fci*-) $Mg_{30}Zn_{60}RE_{10}$ (*RE* = Dy, Er, Gd, Ho, Y), $a(6D) \approx 2 \times 5.2$Å, superspace group $F2/m$-3-5[3, 4] (see Figure 1); (**2**) simple icosahedral (*si*-) Mg-Zn-*RE*: (**2a**) *si*-$Mg_{15}Zn_{75}RE_{10}$, *RE* = Er or Ho: ordered phase[16, 17] and (**2b**) *si*-$Mg_{30+x}Zn_{60}RE_{10-x}$, $x = 0..5$, *RE* = Dy, Gd, Nd, Sm or Y: disordered phase[18]. Both have $a(6D) \approx 5.2$Å and probably belong to superspace group $P2/m$-3-5. Finally a (**3**) decagonal (*d*) phase $d$-$Mg_{40}Zn_{58}RE_2$ (*RE* = Dy, Er, Ho, Lu, Tm and Y) can be characterized by $a(5D) \approx 4.6$Å, $c \approx 5.2$Å[19]. The compositions quoted here are typical values and may differ from case to case by some atom percent, taking into account phase widths.

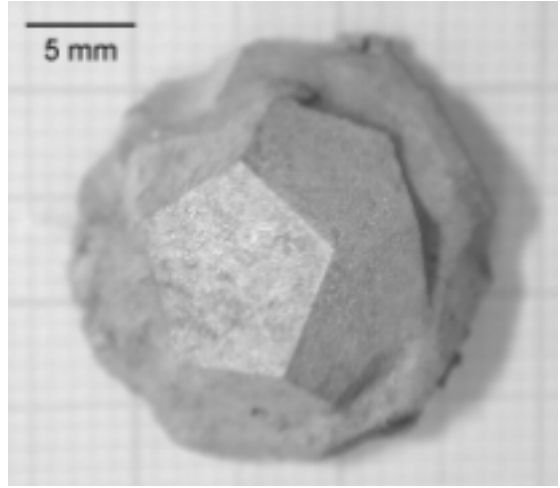

Figure 1: Large single crystal of *fci*-Ho-Mg-Zn (**1**)

*Crystalline Phases*

Some ternary crystalline phases with compositions near the *fci* phase and with known structures today are (**4**) $A$-$Mg_{14}Y_{16}Zn_{70}$ (hP36, $P6_3/mmc$, $a = 8.99$ Å, $c = 9.34$ Å)[20], (**5**) $Z$-$Mg_{28}Y_7Zn_{65}$ (hP92, $P6_3/mmc$, $a = 14.58$ Å, $c = 8.69$ Å)[21], (**6**) $M$-$Mg_{24}Sm_{10}Zn_{65}$ (hP238, $P6_3/mmc$, $a = 23.5$ Å, $c = 8.6$ Å)[22], and (**7**) $\mu_7$-$Mg_{24}Sm_{11}Zn_{65}$ (hP476, $P6_3/mmc$, $a = 33.57$ Å, $c = 8.87$ Å)[23]. (**4**) to (**7**) can be derived from the binary Laves phase $MgZn_2$ (hP12, $P6_3/mmc$, $a = 5.16$ Å, $c = 8.50$ Å)[24] and are considered to be structurally closely related to the *fci* structure. Therefore, sometimes they are called approximants of the icosahedral phase. Interestingly, they do not contain icosahedral clusters as known for *i*-phases and their approximants in other systems.

A cubic phase was found in the Er-Mg-Zn system (**8**): $R$-$Er_{14}Mg_{24}Zn_{62}$ (cF448, $F$-43$m$, $a = 20.02$ Å)[25]. Its structure contains clusters of local symmetry -43$m$ which remind of α-manganese and γ-brass type units.

*Rational approximants*

In order to gain a reasonable structure model for a quasicrystal, the crystal structure determination of a *rational* approximant is a valuable information. It is related to the quasicrystal as follows: To obtain a 3D quasiperiodic structure of icosahedral diffraction symmetry 2/*m*-3-5, an appropriate matrix for cut-and-projection from the 6D hypercubic crystal has to be used[26, 27], *i.e.* it must contain the irrational "golden mean" $\tau = (\sqrt{5} + 1)/2$ in certain matrix elements. A 3D rational Fibonacci $p/q$-approximant of the 6D hypercrystal is generated then using a (rational) ratio of a pair of Fibonacci numbers $p/q = F_{n+1}/F_n$ instead, which approximate τ for $n \to \infty$. Its Laue symmetry necessarily is a subgroup of 2/*m*-3-5, like *m*-3 (cubic), *mmm* (othorhombic) or 2/*m* (monoclinic). The cubic lattice parameter *a*(3D) of a periodic rational approximant is directly connected to the 6D hypercubic lattice constant *a*(6D):

$$a(3D) = 2\, a(6D)\, (p\tau + q) / \sqrt{2 + \tau} \qquad (1)$$

Up to now, no such rational approximants have been observed in the Mg-Zn-*RE* systems. Both space groups $P6_3/mmc$ and $F\text{-}43m$ of above mentioned ternary crystalline phases do not fulfill this requirement.

*Structure data known up to now*

To date, the following detailed structure data are available for *fci*-Mg-Zn-*RE* (**1**): Charrier *et al.* derive from EXAFS measurements for *RE* = Dy, Y that *RE* is essentially surrounded by Zn and a small amount of Mg atoms (12.2 Zn + 3.9 Mg)[28]. Abe and Tsai observe a phase transition to hexagonal Z-$Mg_{28}Y_7Zn_{65}$ (**5**) and therefore conclude that the structural unit of the quasicrystal is not an icosahedral atomic cluster[29]. A single crystal x-ray diffraction study of *fci*-Mg-Y-Zn by Estermann *et al.* reveals no indication of atomic disorder[30]. This is a major characteristic of this class of quasicrytals compared to most of the Al based icosahedral phases. Those often suffer from severe atomic disorder and quasicrystal structure determination therefore is more complicated. Takakura *et al.* extract from their 6D structure analysis of *fci*-Ho-Mg-Zn two dominant short Ho-Ho distances at 5.44Å and 8.80Å[31]. These were corroborated from Z-contrast STEM images.[32] Most recently, after a combined x-ray and HRTEM study, Kramer *et al.* pledge for clusters with 5 shells as a structural element. In *fci*-Mg-Tb-Zn they locate the Tb in an icosahedron of the 2nd shell and in a dodecahedron of the 5th shell[33].

*Classes of* i *quasicrystals*

It is a common systematics to classify icosahedral quasicrystals by distinguishing between Mackay (MI) type and Frank-Kasper (FK) type quasicrystals[34]. The MI type quasicrystal contains Mackay icosahedra as the typical structure element as present in its 1/1-approximant α-Al-Mn-Si (cP138, $Pm\text{-}3$, $a(3D) = 12.68$Å)[35]. This type is characterized by the parameter $a(6D) \approx 4.60$Å and $a/d_A \approx 1.65$; $d_A$ being the mean atomic diameter of the constituents. Parent structure of the FK type quasicrystal is its 1/1-approximant $T\text{-}(Al,Zn)_{49}Mg_{32}$ (cI162, $Im\text{-}3$, $a(3D) = 14.16$Å; Bergman-phase)[36] containing so-called Bergman clusters. For the FK type it is $a(6D) \approx 5.14$Å and $a/d_A \approx 1.75$, so *fci*-Mg-Zn-*RE* (*RE* = Ho: $a(6D) = 5.18$Å; $a/d_A \approx 1.73$) belongs to the latter group. From this point of view, *fci*-Mg-Zn-*RE* should contain Bergman clusters.

*The Frank Kasper concept*

Frank-Kasper phases, on the other hand, do not necessarily contain Bergman clusters. In 1958/59, Frank and Kasper developed an elegant concept for close packings of hard spheres with slightly (about 10%) different radii[37,38]. A lot of theoretical structures result and many of them are adopted by a large number of existing intermetallic phases. They are built up only by slightly distorted tetrahedra and only four "normal" triangulated shells with coordination number (CN) 12, 14, 15 and 16 are possible. The shells are labelled X, R, Q and P, respectively. The atomic environment type (AET) polyhedron code[39] is useful to describe the number of faces (here: triangles) meeting in one vertex[40]: CN12 (X ≡ icosahedron): $12^{5.0}$, CN14 (R): $12^{5.0}2^{6.0}$), CN15 (Q): $12^{5.0}3^{6.0}$) and CN16 (P): $12^{5.0}4^{6.0}$. From all possible arrangements of the four (interpenetrating) polyhedra, two basic concepts emerge to describe the structures: (i) layering of planar or slightly puckered nets and (ii) description *via* "major skeletons". An even more restrictive definition of FK phases is that of so-called tetrahedrally close packed (*tcp*) structures in references 41 and 42. It takes into account the relative number of the "normal" coordinations (P, Q, R, X) for all atoms and certain dihedral angle sum criteria of the coordination polyhedra.

*Outline*

Thus, there is a demand to clarify whether *fci*-Zn-Mg-RE contains icosahedral clusters or not. Another question to answer is whether the FK type quasicrystals are FK phases in the original sense. The short and medium range real structure of *fci*-$Ho_9Mg_{26}Zn_{65}$ will be analyzed to detect possible clusters. Also its relation to the concept of Frank and Kasper will be discussed here.

### III. EXPERIMENTAL AND METHODS

*Phase preparation*

*fci*-Ho-Mg-Zn single crystals were grown from the melt by the liquid-encapsulated top-seeded solution growth method (LETSSG) according to Langsdorf *et al.*[43] 100g of the pure elements (Zn 99.9999%, Mg 99.99% and Ho 99.9%) were put into an aluminia crucible in a molar ratio $n(Ho) : n(Mg) : n(Zn) = 3.0 : 50.5 : 46.5$. To prevent the metals (in particular zinc) from evaporating during crystal growth, 20g of an eutectic KCl/LiCl mixture was added that forms the liquid encapsulation. The crucible was heated under argon up to 973 K where the Ho dissolves in the Mg-Zn melt. A water-cooled tungsten tip was placed into the metallic melt as a nucleation point for the crystal. The temperature was lowered at a rate of 0.6K/h down to 773K. Then the tungsten tip with the crystal was pulled from the melt and the furnace was turned off. The single crystal thus obtained has a diameter of 1.5cm and shows pentagon dodecahedral facetting (Figure 1).

*Phase characterisation*

The composition determined by WDX (Microspec WDX3PC; polished samples, measurement *versus* standard specimen of the pure metals: accuracy ± 1 at.%) is $Ho_9Mg_{26}Zn_{65}$. Parts of a single crystal were used for x-ray precession photographs. They exhibit full Laue symmetry $2/m$-3-5.[17] A portion was crushed and its x-ray powder diffractogram (Siemens Kristalloflex 810, CuKα, λ = 1.541Å) could be indexed with an *fci* lattice parameter $a(6D) = 5.18(3)$Å[17].

*Atomic pair distribution function (PDF)*

Diffractograms of another portion were measured on a Huber Guinier Diffractometer (Seifert system 600) using MoKα$_1$ radiation (λ = 0.70932Å; 2θ = 4..100°; Δθ = 0.01°; $t$ = 60s). High angle data (2θ > 60°) were measured in a second run with triple measuring time. The curves were averaged and background corrected by subtraction of a spline function. Applying the computer program PDFgetX[44], the data were corrected for multiple scattering, polarisation and absorption effects. The structure function $S(Q)$ (Figure 2) was obtained by normalisation and finally the experimental PDF $G(r)_{exp}$ (equation 2a) was calculated.[44]

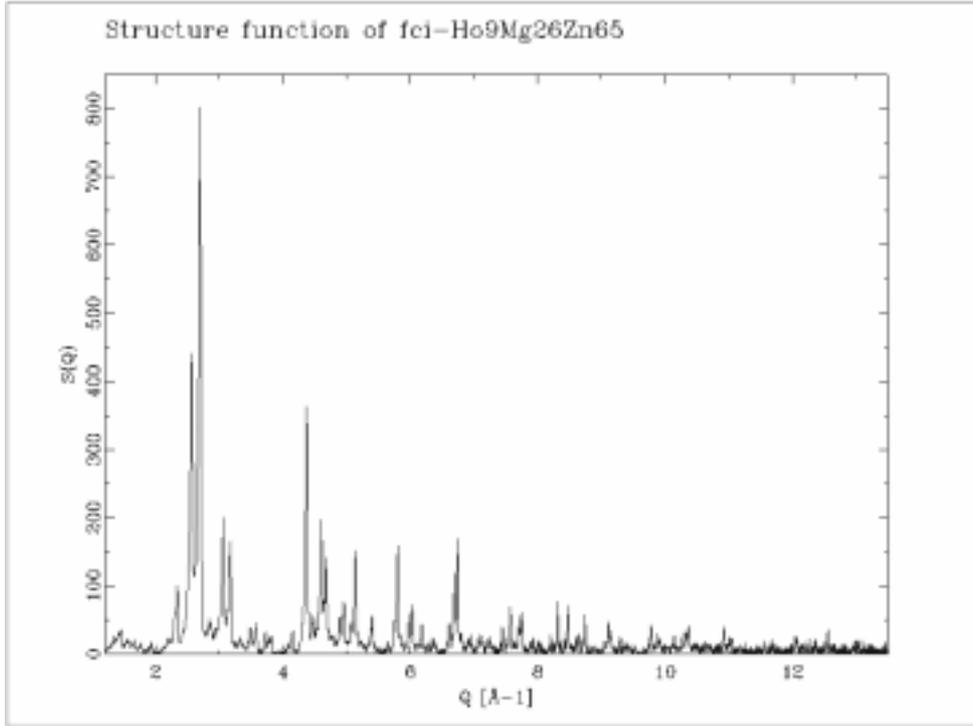

Figure 2.  Structure function $S(Q)$ of $fci$-$Ho_9Mg_{26}Zn_{65}$ from corrected and normalized MoK$\alpha_1$ diffraction data ($2\theta_{max} = 100°$).

The accuracies of experimental PDFs are discussed in ref. 45, it turns out that it depends on the quality of the measurement in high $Q$. A good resolution in $r$ mainly arises from a large $Q_{max} = 4\pi \sin(\theta_{max})/\lambda$. $G(r)$ describes the probability to find any atom (*i.e.* an electron density maximum) i in the distance $r_{ij}$ to any other atom j with respect to an average electron density $\rho_0$ (equation 2). It can be simulated from given structure models using equation (2b); $b_i$ being the scattering factor for atom i and $\langle b \rangle$ is the average scattering factor of the model crystal.

$$G(r) = 4\pi r [\rho(r) - \rho_0] \quad (2)$$

$$G(r)_{exp} = 2/\pi \int_0^\infty Q[S(Q)-1] \sin Q r \, dQ \quad (2a)$$

$$G(r)_{calc} = 1/r \sum_i \sum_j [b_i b_j / \langle b \rangle^2 \, \delta(r-r_{ij})] - 4\pi r \rho_0 \quad (2b)$$

The experimental PDF of the title compound ($r = 0..30$Å, $N = 1000$ data points, $Q_{max} = 13.5$ Å$^{-1}$) was compared to a number of PDFs of known crystal structures calculated using DISCUS and KUPLOT[46]: Using the PDF as a fingerprint of pure materials, it is possible to detect qualitatively differences or similarities between their local and medium range atomic structures.

As will be described in section IV, a suitable starting point for a least squares refinement procedure (PDFFIT[47]) could be found. PDFFIT version 1.2 was compiled to process up to 700 atoms per unit cell and implemented on a personal computer. The only parameters fitting the shape of the curve applied by PDFFIT are the scale factor and the dynamic correlation factor $\delta$ which accounts for correlated motion. Structure parameters are a (virtual) lattice constant $a'$ (3D) to define a coordinate system, atomic positions, temperature- and occupation factors. The refinement is performed in a confined $r$ range which describes the quasicrystal structure locally: $r_{max} \approx 1.2 \times a'(3D)$. It terminates when a minimum of $R$ (equation 3) is reached. During the process a relaxation factor has to be set in order to meet the *global* minimum.

$$R = \sqrt{\{\sum_{i=1}^{N}[G_{obs}(r_i)-G_{calc}(r_i)]^2 / \sum_{i=1}^{N} G^2_{obs}(r_i)\}} \qquad (3)$$

Reasonable PDF structure models may have $R = 0.10$ to $0.35$ and cannot be compared to conventional Rietveld refinements. The question whether the $R$ value is "good" or not is *e.g.* discussed in ref. 48. In our experience up to now, $R$ is a function of a variety of parameters, such as the resolution of the PDF, $Q_{max}$; number of data points, $N$; $r$ range or the number of parameters. Nevertheless, its evolution leads to a minimum, *i.e.* it shows the best fit, all the above parameters kept constant. A visual inspection then of original, fitted and difference plots for $G(r)$ in one graphic is more informative and easily done (Figures 5 and 6). – Anyway, once the refinement has converged, the critical test should be whether the resulting structural features are chemically and physically sound.

## IV. STRUCTURE MODEL CONSTRUCTION

*PDF fingerprints*

To get a good starting model of *fci*-$Ho_9Mg_{26}Zn_{65}$ for consecutive structure refinements *via* its PDF, in a first step PDFs of the related ternary phases described in section II were simulated. They were compared visually in the range $r = 0..30$Å to the experimental PDF of the quasicrystal. None of the hexagonal phases (**4**) to (**7**) nor the cubic *R*-phase (**8**) shows convincing overall similarities though limited features of the PDFs may be reproduced. This is also true for Laves' $MgZn_2$ which is sometimes referred to as the "master compound" of *i*-Mg-Zn-*RE* quasicrystals (Figure 3a). In the binary system Mg-Zn two other compounds can be considered as rational approximants: $Mg_2Zn_{11}$ (cP39, *Pm*-3, a = 8.552Å: 1/0-approximant)[49] and the pseudocubic 1/1-1/1-1/1-approximant $Mg_{51}Zn_{20}$ (oI158, *Immm*, $a = 14.083$Å, $b = 14.486$Å, $c = 14.025$Å)[50]. $Mg_2Zn_{11}$ is built up of the inner core of the Bergman cluster only (44 atoms in two shells; so-called Pauling triacontahedron). $Mg_{51}Zn_{20}$ was shown to contain distorted Mackay type clusters[51]. Again, only poor similarities could be observed.

In the pseudo binary systems Mg-(Zn,*B*) (*B*=Al, Ga), there is a series of large approximant structures known: 1/1-Al-Mg-Zn (re-determined Bergman phase: cI160, *Im*-3, $a = 14.217$Å)[52], 2/1-Al-Mg-Zn (cP676, *Pa*-3, $a = 23.034$Å)[52, 53], 3/2-2/1-2/1-Ga-Mg-Zn (oC1104, *Cmc*$2_1$, $a = 36.840$Å $b = 22.782$Å $c = 22.931$Å)[54]. Interestingly, 1/1-Al-Mg-Zn yields a global qualitative match to the PDF of *fci*-Ho-Mg-Zn. This 1/1-approximant mirrors almost all the maxima and minima in the pattern up to $r = 30$Å (Figure 3b), the similarity is further improved in the comparison to 2/1-Al-Mg-Zn.

*Bergman cluster decorated CCT*

Those approximants are all built of Bergman clusters which are arranged according to Henleys canonical cell tiling (CCT)[55]: The CCT recently was decorated by Kreiner[54] to describe and solve the crystal structures. A Bergman cluster consists of 104 atoms which are arranged in three shells and is built starting from an empty center position, an orbit called $\alpha^0$. The first shell of 12 atoms is the inner icosahedron $\alpha^1$. The second shell is formed by (a) 20 atoms building a pentagon dodecahedron $\beta$, and (b) 12 further atoms (outer icosahedron $\alpha^2$). The cluster is completed by 60 atoms forming a soccer ball ($\alpha^3$ and $\gamma$) in the third shell. Clusters can be linked along their 2fold or 3fold axes, sharing edges or hexagonal faces, the links being called "*b*-bonds" or "*c*-bonds" in the CCT, respectively.[55] The remaining space between the clusters is filled up by a number of glue atoms (δ, ε and η).

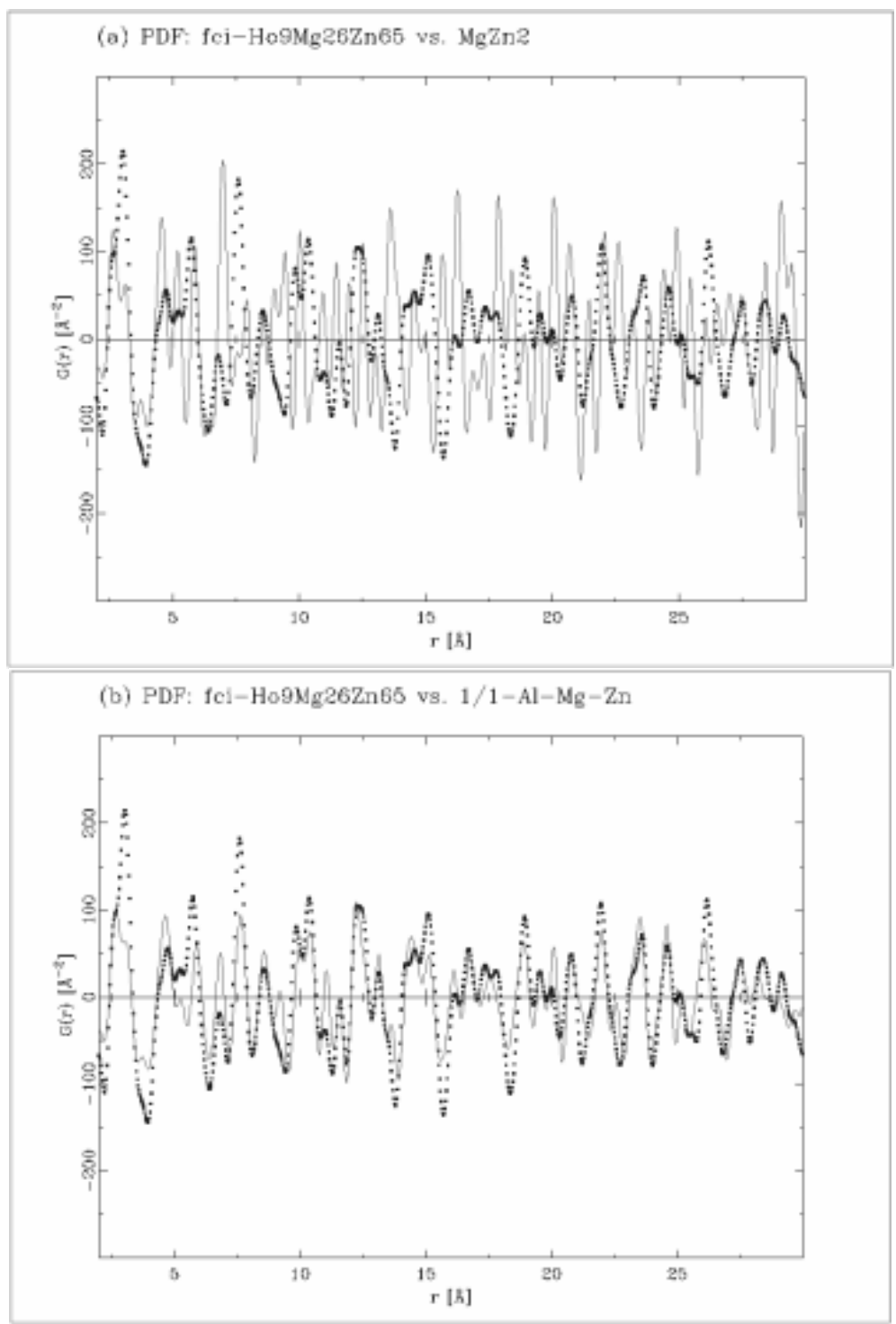

Figure 3. Qualitative comparison of $G(r)_{exp}$ of *fci*-$Ho_9Mg_{26}Zn_{65}$ (dots) to (a) $G(r)_{calc}$ $MgZn_2$ (solid line) and (b) $G(r)_{calc}$ 1/1-Al-Mg-Zn (solid line).

*Model construction*

The striking similarity of the PDFs in Figure 3b was encouraging to construct a "virtual" 1/1-approximant for *fci*-Ho-Mg-Zn. Even though there are no existing rational approximants of *fci*-Zn-Mg-*RE*, the quasicrystal may be related to hypothetical ("virtual") rational approximants.

The average intermetallic diameters[56] $d_{av}$ of alumnium and zinc are almost equal. It is $d_{av}(Al) = 2.864$Å $\approx d_{av}(Zn) = 2.788$Å, in 1/1-Al-Mg-Zn they substitute for each other on the same crystallographic sites with coordination number (CN) 11 or 12. On the other hand, holmium is larger than magnesium but still of comparable size: $d_{av}(Ho) = 3.532$Å $\approx d_{av}(Mg) = 3.204$Å. In 1/1-Al-Mg-Zn, Mg has CN14, 15 and 16. In ref. 28 CN 16 has been calculated for the *RE* atoms in the quasicrystal. Thus, in *fci*-Ho-Zn-Mg, Ho might be present on Mg sites with CN16.

To build a "virtual" 1/1-approximant model from 1/1-Al-Mg-Zn, all Al atoms were replaced by Zn. Mg with CN16 is located on two crystallographic sites (Mg1 on 24*g* and Mg2 on 16*f*). Together, both β-sites form a pentagon dodecahedron in the second shell of the Bergman cluster. If we replace Mg2 by holmium, the hypothetic formula $Ho_{10}Mg_{30}Zn_{60}$ results (Table 1). Indeed, $Mg_{30}Zn_{60}RE_{10}$ is the typical average composition of the *fci* phase in the Mg-Zn-*RE* systems. The Ho atoms then form a cube inserted in the shell 2a. The edge length of this cube is approximately 5.2Å and it is completed to the pentagon dodecahedron by 12 Mg atoms. This way no direct *intra*-cluster Ho-Ho contacts are generated which in fact should not be present according to EXAFS data[28].

For reasons of better comparability, the atom labelling of the Al-Mg-Zn phases is retained where possible. Using equation (1) and the observed quasilattice constant of the Ho compound, the lattice constant of the "virtual" 1/1-approximant gives $a'_{"1/1"}(3D) = 14.26$Å. As a drawback, the "1/1" model still suffers from direct Ho-Ho contacts of 3.1Å. There are 1.0 such *inter*-cluster contacts per average Ho atom.

| atom in 1/1-$Mg_{32}(Al,Zn)_{48}$ | atom in "1/1"-$Ho_{10}Mg_{30}Zn_{60}$ | Wyckoff position | x/a | y/a | z/a | CN | shell | structural function |
|---|---|---|---|---|---|---|---|---|
| *M*1 | Zn1 | 48*h* | 0.09659 | 0.30933 | 0.34228 | 12 | 3. | soccer ball |
| Mg1 | Mg1 | 24*g* | 0 | 0.1169 | 0.3004 | 16 | 2a. | pentagon dodecahedron |
| *M*2 | Zn2 | 24*g* | 0 | 0.15117 | 0.09283 | 11 | 1. | inner icosahedron |
| *M*3 | Zn3 | 24*g* | 0 | 0.30674 | 0.17925 | 12 | 2b. | outer icosahedron |
| Mg2 | Ho1 | 16*f* | 0.18610 | *x* | *x* | 16 | 2a. | pentagon dodecahedron/cube |
| Mg3 | Mg3 | 12*e* | 0.0974 | 0 | ½ | 14 | 3. | soccer ball |
| Mg4 | Mg4 | 12*e* | 0.30192 | 0 | ½ | 15 | — | glue atom |

Table 1. Crystal data for 1/1-$Mg_{32}(Al,Zn)_{48}$ cited from ref. 52 for comparison and the starting model "1/1"-$Ho_{10}Mg_{30}Zn_{60}$ used in this work: cI160, *Im*-3, $a \approx 14$Å; *M* = (Al,Zn)

Analogously, 2/1-Al-Mg-Zn can be altered into a hypothetical "2/1"-Ho$_9$Mg$_{34}$Zn$_{57}$. Here, direct Ho-Ho contacts can be almost avoided by proper choice of the Mg atoms to be replaced by Ho atoms. The Ho cubes then are tilted with respect to the cubes in neighbouring clusters. Per average Ho atom there are only 0.125 direct contacts Ho-Ho left. In the "2/1"-Ho$_9$Mg$_{34}$Zn$_{57}$ model four of the five possible orientations of a cube inscribed in a pentagon dodecahedron (Figure 4) as permitted by the symmetry operations of space group *Pa*-3, are implemented. We assume that in the quasicrystal all five possible orientations of the cubes as part of the pentagon dodecahedron are realized. The full icosahedral symmetry 2/*m*-3-5 could be retained this way; see Figure 4.

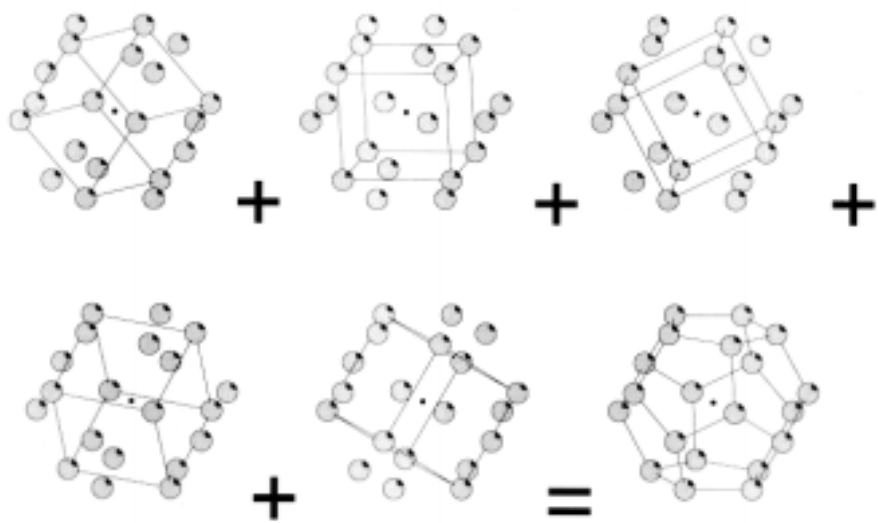

Figure 4.    20 atoms form the vertices of a pentagon dodecahedron. A subset of 8 atoms can be arranged on the vertices of a cube. The cube edge length is τ times the dodecahedron edge length. Five different orientations of the cube are possible.

In ref. 52 two possible variants are given, in which either the atom position *M*2 or *M*4 (see Table 2) is occupied. We use the "*M*2-variant" which is the geometrically more ideal case given by the decorated CCT. *M*20 was omitted in the start model because it is only partially occupied in 2/1-Al-Mg-Zn. Table 2 comprises the data of 2/1-Al-Mg-Zn and the virtual "2/1"-Ho-Mg-Zn approximant. Again, atom labelling is retained where possible.

| atom in 2/1-$Al_{93}Mg_{290}Zn_{293}$ | atom in "2/1"-$Ho_9Mg_{34}Zn_{57}$ | Wyckoff position | x/a | y/a | z/a | CN | shell | structural function |
|---|---|---|---|---|---|---|---|---|
| Mg1 | Mg1 | 24d | 0.0268 | 0.3496 | 0.4492 | 15 | — | glue atom |
| M1 | Zn1 | 24d | 0.03111 | 0.46474 | 0.15763 | 12 | 2b. | outer icosahedron |
| Mg2 | Ho1 | 24d | 0.03320 | 0.3458 | 0.22464 | 16 | 2a. | pentagon dodecahedron/cube |
| Mg3 | Mg3 | 24d | 0.03373 | 0.3446 | 0.08322 | 16 | 2a. | pentagon dodecahedron |
| M2 | Zn2 | 24d | 0.03482 | 0.13143 | 0.09420 | 12 | | soccer ball |
| M3 | Zn3 | 24d | 0.03482 | 0.22329 | 0.15033 | 12 | 2b. | outer icosahedron |
| Mg4 | Mg4 | 24d | 0.03887 | 0.1585 | 0.3514 | 15 | — | glue atom |
| M4 | — | 24d | 0.0423 | 0.0800 | 0.1379 | 12 | — | glue atom |
| M´5 | Zn5 | 24d | 0.05612 | 0.12449 | 0.23284 | 14 | 3. | soccer ball, edge connecting |
| M6 | Zn6 | 24d | 0.05927 | 0.28774 | 0.34327 | 12 | 3. | soccer ball |
| M7 | Zn7 | 24d | 0.05946 | 0.09386 | 0.46418 | 12 | 3. | soccer ball |
| M8 | Zn8 | 24d | 0.06132 | 0.40346 | 0.34251 | 12 | 3. | soccer ball |
| Mg5 | Mg5 | 24d | 0.0778 | 0.2225 | 0.4557 | 14 | 3. | soccer ball, edge connecting |
| M´9 | Zn9 | 24d | 0.09806 | 0.46149 | 0.44460 | 12 | 3. | soccer ball |
| M10 | Zn10 | 24d | 0.09837 | 0.46253 | 0.25006 | 12 | 3. | soccer ball |
| M´11 | Zn11 | 24d | 0.09925 | 0.23121 | 0.25143 | 12 | 3. | soccer ball |
| M12 | Zn12 | 24d | 0.13184 | 0.40555 | 0.15555 | 12 | 3. | soccer ball |
| M´13 | Zn13 | 24d | 0.13456 | 0.28894 | 0.15842 | 12 | 3. | soccer ball |
| Mg6 | Ho2 | 24d | 0.1579 | 0.3469 | 0.41621 | 16 | 2a. | pentagon dodecahedron/cube |
| M14 | Zn14 | 24d | 0.15890 | 0.23335 | 0.35314 | 12 | 2b. | outer icosahedron |
| Mg7 | Mg7 | 24d | 0.16139 | 0.3478 | 0.27503 | 16 | 2a. | pentagon dodecahedron |
| M15 | Zn15 | 24d | 0.16161 | 0.46035 | 0.34761 | 12 | 2b. | outer icosahedron |
| Mg8 | Mg8 | 24d | 0.2301 | 0.2324 | 0.4637 | 16 | 2a. | pentagon dodecahedron |
| Mg9 | Mg9 | 24d | 0.2302 | 0.4640 | 0.4611 | 16 | 2a. | pentagon dodecahedron |
| M16 | Zn16 | 24d | 0.25300 | 0.29003 | 0.34736 | 11 | 1. | inner icosahedron |
| M17 | Zn17 | 24d | 0.25506 | 0.40467 | 0.34554 | 11 | 1. | inner icosahedron |
| M18 | Zn18 | 24d | 0.28631 | 0.34802 | 0.44164 | 11 | 1. | inner icosahedron |
| M19 | Zn19 | 24d | 0.34656 | 0.43980 | 0.40367 | 11 | 1. | inner icosahedron |
| M20 | — | 8c | 0.0293 | x | x | 12 | — | glue atom |
| Mg10 | Mg10 | 8c | 0.15516 | x | x | 16 | 2a. | glue atom |
| Mg11 | Ho3 | 8c | 0.23336 | x | x | 16 | 2a. | pentagon dodecahedron/cube |
| Mg12 | Ho4 | 8c | 0.4606 | x | x | 16 | 2a. | pentagon dodecahedron/cube |

Table 2. Crystal data for 2/1-$Al_{93}Mg_{290}Zn_{293}$ cited from ref. 52 for comparison and the starting model "2/1"-$Ho_9Mg_{34}Zn_{57}$ used in this work: cP676, Pa-3, a ≈ 23Å; M = (Al,Zn), M´ = (Mg,Zn)

# V. STRUCTURE REFINEMENTS

*"1/1" model*

The "1/1"-Ho$_{10}$Mg$_{30}$Zn$_{60}$ model with the positional parameters of the 1/1-Al-Mg-Zn phase as given in Table 1 was used to start. First of all, only the scale factor and the dynamic correlation factor δ in a limited range $r = 2..10$Å were refined, the $R$ value dropping to 0.5. Extension to $r = 20$Å and refining of the virtual lattice constant and an overall temperature factor leads to $R = 0.4$. Allowing individual temperature factors for symmetry equivalent atoms (note: *as if* in *Im*-3) converges at $R = 0.28$. To check, whether the "cube arrangement" of Ho1 atoms is realistic, Mg and Ho was statistically distributed on the pentagon dodecahedron sites Mg1 and Mg2 (initial occupation factors for each site: o[Mg] = 0.6, o[Ho] = (1 - o[Mg]) = 0.4 which results in the same composition). All temperature factors were fixed to $U_{eq} = 0.015$Å$^2$ and the other occupation factors were refined, too. The refinement shifts the occupations back to o[Mg] = 0.9 for site Mg1 and to o[Mg] = 0.4 for Mg2. At the same time, the occupation for the Mg4 site was increased about factor 5, so that some Ho might be present as glue atom. Convergence at $R = 0.25$ indicates that the "cube arrangement" seems to be reasonable; it is restored for the further refinements. A Ho2 atom with the same positional parameters of Mg4 is introduced, initial occupation factors are o[Ho2] = (1 - o[Mg4]) = 0.25. The refinement shifts o[Ho2] to 0.265 and again converged at $R = 0.25$. Finally, all positional parameters (also constrained *as if* in *Im*-3) were allowed to refine. The $R$ value drops to 0.23. In a further test, an additional Zn atom was laid in the center of the cluster (2*a*: 000), as published for the *T* phase in 1957[36], but this is not appreciated in the refinement ($R = 0.24$, o[Zn0] = 0.7).

If one assumes that the local structure of the quasicrystal is present in, say, a sphere with a diameter about 120% of an approximants lattice constant, reasonable ranges would be $r_{max} = 1.2 \times 14$Å $\approx 17$Å for "1/1"-Ho-Mg-Zn and $r_{max} = 1.2 \times 23$Å $\approx 27$Å for "2/1"-Ho-Mg-Zn. So the final data of "1/1"-Ho-Mg-Zn are calculated for $r = 2..17$Å yielding $R = 0.218$. *fci*-Ho-Mg-Zn, thougt as locally arranged like an 1/1-approximant would then have the composition Ho$_{12.0}$Mg$_{28.0}$Zn$_{60.0}$; compare to the WDX data Ho$_9$Mg$_{26}$Zn$_{65}$. Final atom positions are used to calculate bond legths and coordination geometries using DIAMOND[57] and are summarized in Table 3. Figure 5 plots measured, calculated and difference PDFs. Details of the final refinement procedure are given in Table 5.

| atom | $x$ | $y$ | $z$ | $U_{eq}/$ 10$^{-2}$ Å$^2$ | $d_{min}$/Å | $d_{max}$/Å | <$d$>/Å | CN | polyhedron code[40] (FK) | structural function |
|---|---|---|---|---|---|---|---|---|---|---|
| Zn1 | 0.0975 | 0.3088 | 0.3427 | 2.4 | 2.606 | 3.226 | 2.934 | 12 | 12$^{5.0}$ (X) | α$^3$ - soccer ball |
| Mg1 | 0 | 0.128 | 0.284 | 1.0 | 2.742 | 3.773 | 3.182 | 16 | 12$^{5.0}$4$^{6.0}$ (P) | β - pentagon dodeca-hedron |
| Zn2 | 0 | 0.1512 | 0.0933 | 0.6 | 2.531 | 3.018 | 2.765 | 11 | 7$^{5.0}$5$^{3.0.1}$ | α$^1$ - inner icosahedron |
| Zn3 | 0 | 0.3103 | 0.1721 | 6.7 | 2.531 | 3.332 | 2.933 | 12 | 12$^{5.0}$ (X) | α$^2$ - outer icosahedron |
| Ho1 | 0.1867 | $x$ | $x$ | 3.0 | 3.018 | 3.190 | 3.101 | 16 | 12$^{5.0}$4$^{6.0}$ (P) | β - pentagon dodeca-hedron/cube |
| Mg3 | 0.085 | 0 | ½ | 2.5 | **2.414** | 3.773 | 3.077 | 14 | 12$^{5.0}$2$^{6.0}$ (R) | γ - soccer ball |
| Mg4/ Ho2 | 0.3097 | 0 | ½ | 1.5 | 2.963 | 3.332 | 3.135 | 15 | 12$^{5.0}$3$^{6.0}$ (Q) | δ - glue atom |
| [void] | 0 | 0 | 0 | [—] | 2.518 | 2.518 | **2.518** | [12] | [12$^{5.0}$ (X)] | α$^0$ - cluster center |

Table 3.  Structural parameters of *fci*- Ho$_9$Mg$_{26}$Zn$_{65}$ refined as a virtual approximant "1/1"-Ho$_{12.0}$Mg$_{28.0}$Zn$_{60.0}$ *as if* in *Im*-3 ($R = 21.8\%$). $x$, $y$ and $z$ are fractions of $a´ = 14.22$Å; $d_{min}$ ($d_{max}$; <$d$>) is the minimal (maximal; average) distance of an atom from its coordinating neighbours. Note the strikingly short bond lengths in **bold**. It is o[Mg4] = (1 - o[Ho2]) = 0.735.

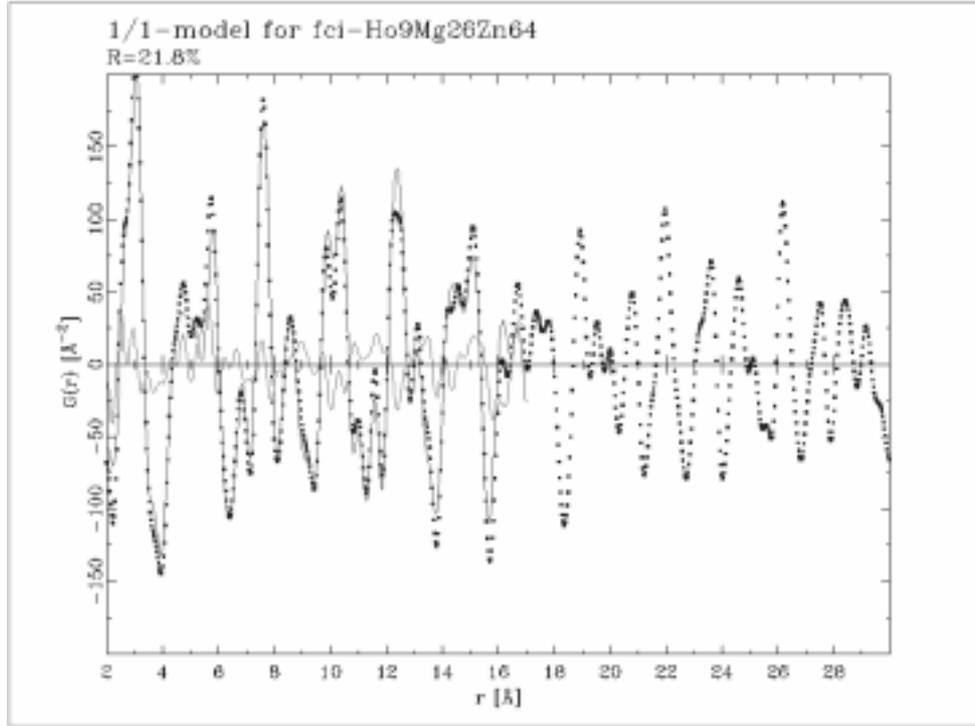

Figure 5. PDF from measured diffraction data of *fci*- $Ho_9Mg_{26}Zn_{65}$ (dots), PDF calculated from virtual approximant "1/1"-$Ho_{12.0}Mg_{28.0}Zn_{60.0}$ ($r_{max}$ = 17Å, solid line) and their difference plot; $R$ = 21.8% (grey line).

*"2/1" model*

Structure data of 2/1-Al-Mg-Zn as given in Table 2 were used as a starting point for a second refinement series. After a quite reasoanble scale factor was found ($R$ = 0.38), the model was allowed to refine the virtual lattice parameter and temperature factors. The latter were treated *as if* constrained by the symmetry operations of *Pa*-3 for each orbit of atoms. Cycles calculating $G(r = 2..20Å)$ converge at $R$ = 0.25. As the temperature factor $U_{eq}[Mg10]$ became negative, Mg10 was replaced by Ho5, leading to an $R$ value of 0.19. The $r$ range could then be extended to 27Å, resulting in $R$ = 0.20.

Subsequently, the positional parameters were constrained *as if* in *Pa*-3 and allowed to refine with a large relaxation factor applied. The refinement smoothly converged to $R$ = 0.14. Then Zn20 was added on the position of *M*20 which was appreciated by the refinement: $R$ = 0.13. On the other hand $R$ is increased about 0.01 on inserting another Zn atom at the cluster center (site 8*c*: *xxx*; $x$ = 0.346); it was removed again. The final structural parameters of "2/1"-Ho-Mg-Zn are calculated for $r$ = 2..27Å, converging at $R$ = 0.129.

Thus, the local structure of *fci*-Ho-Mg-Zn up to almost 30Å can be well described as arranged like in an imaginary 2/1-approximant having the composition $Ho_{10.6}Mg_{24.7}Zn_{64.7}$. This fits the WDX data $Ho_9Mg_{26}Zn_{65}$ quite well. Final atom positions, bond lengths and coordination geometries are summarized in Table 4. Figure 6 plots measured, calculated and difference PDFs. Details of the final refinement procedure are given in Table5.

| atom | $x$ | $y$ | $z$ | $U_{eq}/$ $10^{-2}$ Å$^2$ | $d_{min}$/Å | $d_{max}$/Å | $<d>$/Å | CN | polyhedron code[40] (FK) | structural function |
|---|---|---|---|---|---|---|---|---|---|---|
| Mg1 | 0.0352 | 0.3518 | 0.4563 | 1.0 | 2.925 | 3.617 | 3.134 | 15 | $12^{5.0}3^{6.0}$ (Q) | $\delta^X$ - glue atom |
| Zn1 | 0.0294 | 0.4600 | 0.1629 | 3.8 | 2.446 | 4.056 | 3.031 | 12 | $12^{5.0}$ (X) | $\alpha^2$ - outer icosahedron |
| Ho1 | 0.0330 | 0.3486 | 0.2267 | 1.4 | 2.917 | 3.745 | 3.106 | 16 | $12^{5.0}4^{6.0}$ (P) | $\beta$ - pentagon dodecahedron/cube |
| Zn2 | 0.0379 | 0.1303 | 0.0913 | 3.1 | 2.550 | 3.340 | 2.916 | 12 | $12^{5.0}$ (X) | $\alpha^3$ - soccer ball |
| Mg3 | 0.0523 | 0.3120 | 0.0697 | 1.0 | 3.155 | 3.855 | 3.270 | 16 | $12^{5.0}4^{6.0}$ (P) | $\beta$ - pentagon dodecahedron |
| Zn3 | 0.0298 | 0.2146 | 0.1658 | **11.7** | 2.569 | 3.373 | 2.996 | 12 | $12^{5.0}$ (X) | $\alpha^2$ - outer icosahedron |
| Mg4 | 0.0346 | 0.1592 | 0.3477 | 1.0 | **2.460** | 4.468 | 3.134 | 15 | $12^{5.0}3^{6.0}$ (Q) | $\delta^X$ - glue atom |
| Zn5 | 0.0350 | 0.1022 | 0.2353 | **12.4** | **2.214** | 3.686 | 3.072 | 14 | $12^{5.0}2^{6.0}$ (R) | $\gamma$ - soccer ball, edge connecting |
| Zn6 | 0.0560 | 0.2896 | 0.3409 | 0.9 | 2.524 | 3.605 | 2.908 | 12 | $12^{5.0}$ (X) | $\alpha^3$ - soccer ball |
| Zn7 | 0.0546 | 0.0910 | 0.4603 | 2.7 | 2.603 | 3.560 | 2.965 | 12 | $12^{5.0}$ (X) | $\alpha^3$ - soccer ball |
| Zn8 | 0.0654 | 0.4017 | 0.3430 | 2.4 | 2.596 | 3.098 | 2.847 | 12 | $12^{5.0}$ (X) | $\alpha^3$ - soccer ball |
| Mg5 | 0.0909 | 0.2061 | 0.4717 | 2.7 | **2.214** | 4.185 | 3.176 | 14 | $12^{5.0}2^{6.0}$ (R) | $\gamma$ - soccer ball, edge connecting |
| Zn9 | 0.0978 | 0.4601 | 0.4462 | 0.6 | 2.550 | 3.364 | 2.954 | 12 | $12^{5.0}$ (X) | $\alpha^3$ - soccer ball |
| Zn10 | 0.0939 | 0.4590 | 0.2508 | 1.8 | 2.534 | 3.608 | 2.990 | 12 | $12^{5.0}$ (X) | $\alpha^3$ - soccer ball |
| Zn11 | 0.0963 | 0.2266 | 0.2606 | 2.8 | 2.524 | 3.241 | 2.897 | 12 | $12^{5.0}$ (X) | $\alpha^3$ - soccer ball |
| Zn12 | 0.1144 | 0.3957 | 0.1440 | 3.0 | 2.444 | 3.884 | 2.971 | 12 | $12^{5.0}$ (X) | $\alpha^3$ - soccer ball |
| Zn13 | 0.1350 | 0.2938 | 0.1654 | 2.3 | 2.444 | 3.275 | 2.933 | 12 | $12^{5.0}$ (X) | $\alpha^3$ - soccer ball |
| Ho2 | 0.1571 | 0.3458 | 0.4152 | 5.4 | 2.960 | 3.689 | 3.172 | 16 | $12^{5.0}4^{6.0}$ (P) | $\beta$ - pentagon dodecahedron/cube |
| Zn14 | 0.1615 | 0.2316 | 0.3568 | 3.7 | 2.526 | 3.363 | 2.989 | 12 | $12^{5.0}$ (X) | $\alpha^2$ - outer icosahedron |
| Mg7 | 0.1550 | 0.3458 | 0.2793 | 1.1 | 2.862 | 3.791 | 3.165 | 16 | $12^{5.0}4^{6.0}$ (P) | $\beta$ - pentagon dodecahedron |
| Zn15 | 0.1591 | 0.4649 | 0.3493 | 5.7 | 2.541 | 3.276 | 2.843 | 12 | $12^{5.0}$ (X) | $\alpha^2$ - outer icosahedron |
| Mg8 | 0.2295 | 0.2454 | 0.4853 | 9.8 | 2.639 | 3.791 | 3.153 | 16 | $12^{5.0}4^{6.0}$ (P) | $\beta$ - pentagon dodecahedron |
| Mg9 | 0.2168 | 0.4589 | 0.4485 | 2.7 | 2.649 | 4.185 | 3.180 | 16 | $12^{5.0}4^{6.0}$ (P) | $\beta$ - pentagon dodecahedron |
| Zn16 | 0.2531 | 0.2898 | 0.3475 | 0.8 | 2.526 | 3.397 | 2.888 | 11 | $7^{5.0}5^{3.0}1$ | $\alpha^1$ - inner icosahedron |
| Zn17 | 0.2550 | 0.4046 | 0.3455 | 0.2 | 2.593 | 3.241 | 2.828 | 11 | $7^{5.0}5^{3.0}1$ | $\alpha^1$ - inner icosahedron |
| Zn18 | 0.2911 | 0.3502 | 0.4443 | 1.2 | 2.597 | 3.159 | 2.862 | 11 | $7^{5.0}5^{3.0}1$ | $\alpha^1$ - inner icosahedron |
| Zn19 | 0.3466 | 0.4397 | 0.4037 | 0.8 | 2.446 | 3.855 | 2.917 | 11 | $7^{5.0}5^{3.0}1$ | $\alpha^1$ - inner icosahedron |
| Zn20 | 0.0277 | $x$ | $x$ | 3.3 | **2.210** | 3.354 | 2.917 | 13 | $1^{3.0}0^{9}5^{5.0}3^{6.0}$ | $\delta^Z$ - glue atom |
| Ho5 | 0.1546 | $x$ | $x$ | 0.4 | 2.836 | 3.341 | 3.198 | 16 | $12^{5.0}4^{6.0}$ (P) | $\delta^Y$ - glue atom |
| Ho3 | 0.2267 | $x$ | $x$ | **15.2** | 2.836 | 3436 | 3.191 | 16 | $12^{5.0}4^{6.0}$ (P) | $\beta$ - pentagon dodecahedron/cube |
| Ho4 | 0.4607 | $x$ | $x$ | 0.3 | 2.976 | 4.071 | 3.264 | 16 | $12^{5.0}4^{6.0}$ (P) | $\beta$ - pentagon dodecahedron/cube |
| [void] | 0.3458 | $x$ | $x$ | [—] | **2.491** | 2.595 | **2.535** | [12] | [$12^{5.0}$ (X)] | $\alpha^0$ - cluster center |

Table 4. Structural parameters of *fci*- Ho$_9$Mg$_{26}$Zn$_{65}$ refined as a virtual approximant "2/1"-Ho$_{10.6}$Mg$_{24.7}$Zn$_{64.7}$ *as if* in *Pa*-3 ($R$ = 12.9%). $x$, $y$ and $z$ are fractions of $a'$ = 23.03 Å; $d_{min}$ ($d_{max}$; $<d>$) is the minimal (maximal; average) distance of an atom from its coordinating neighbours. Note the strikingly short bond lengths or large temperature factors in **bold**.

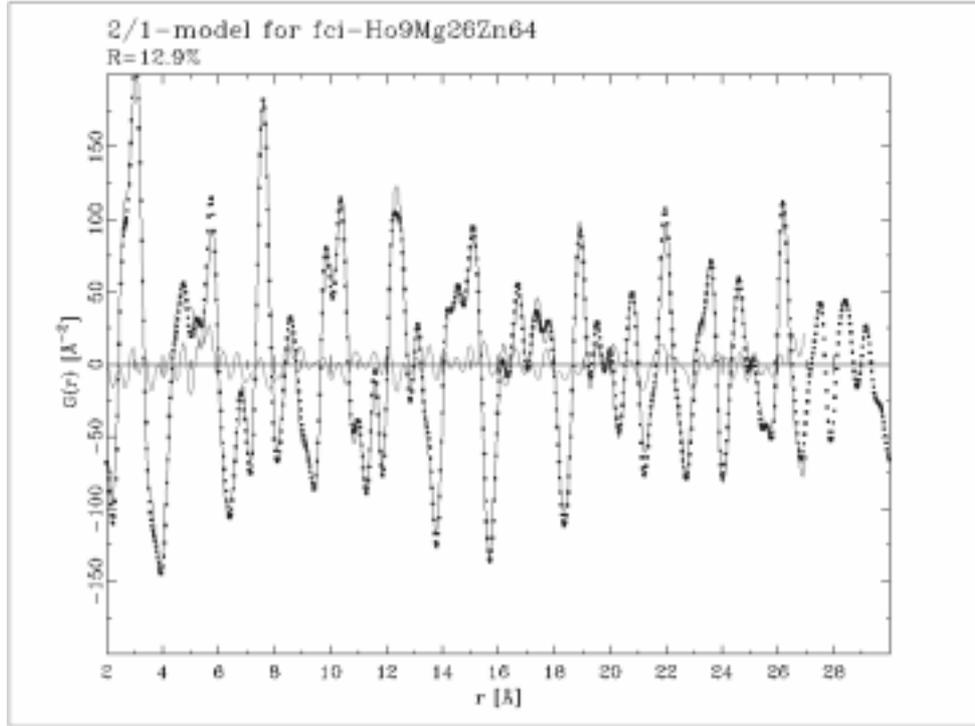

Figure 6.     PDF from measured diffraction data of *fci*- $Ho_9Mg_{26}Zn_{65}$ (dots), PDF calculated from virtual approximant "2/1"-$Ho_{10.6}Mg_{24.7}Zn_{64.7}$ ($r_{max}$ = 27Å, solid line) and their difference plot; $R$ = 12.9% (grey line).

| refinement data | "1/1"-$Ho_{12.0}Mg_{28.0}Zn_{60.0}$ | "2/1"-$Ho_{10.6}Mg_{24.7}Zn_{64.7}$ |
|---|---|---|
| scale factor | 48.081(19) | 58.749(7) |
| dynamic correlation factor $\delta$ | 0.5247(16) | 0.677906(9) |
| low $r/\sigma$ ratio | 1.0 | 1.0 |
| virtual approximant space group (ref. 58) | $Im$-3 (no. 202) | $Pa$-3 (no. 205) |
| virtual approximant lattice parameter $a'$(3D) /Å | 14.21739(12) | 23.03320(3) |
| hypercubic lattice parameter $a$(6D) /Å (from eq.1) | 5.165 | 5.171 |
| data range in $r$ /Å | 0.. 30 | 0.. 30 |
| calculated $r$ range in Å | 0.. 19.9630 | 0.. 29.9630 |
| refinement $r$ range in Å | 2.. 17 | 2.. 27 |
| number of data points used | 499 | 832 |
| radiation | MoK$\alpha_1$ | MoK$\alpha_1$ |
| $\lambda$/Å | 0.70932 | 0.70932 |
| termination at $Q_{max}$ /Å$^{-1}$ | 13.5 | 13.5 |
| $Q$ resolution $\sigma(Q)$ /Å$^{-1}$ | 0.01 | 0.01 |
| number of refined parameters | 23 | 120 |
| magic number (relaxation factor) | 21.5 | 36.0 |
| $R$-values | 0.21840942 | 0.12910655 |
| change last cycle | -0.00091560 | -0.00000143 |
| correlations greater than 0.8 | none | none |

Table 5.     Data of the final PDFFIT refinements of *fci*- $Ho_9Mg_{26}Zn_{65}$ as virtual "1/1"-$Ho_{12.0}Mg_{28.0}Zn_{60.0}$ and "2/1"-$Ho_{10.6}Mg_{24.7}Zn_{64.7}$, respectively.

## VI. DISCUSSION

First of all, it has to be kept in mind that our local or medium range model PDF refinements are not based on the hard constraint of translational symmetry. Though, the PDFs of the models were refined using cubic translational cells as coordinate systems, containing the symmetry elements of space groups *Im*-3 and *Pa*-3, respectively. This allows the local icosahedral symmetry to be described in a comfortable way. Nevertheless, simulated powder diffractograms from the above models exhibit close qualitative resemblance to the experimental diffraction pattern. But they are not refinable by the conventional Rietveld method as the positions of the reflections given by a 3D cubic lattice cannot be (and *are* not) the same as from the 6D hypercubic lattice found for *fci*-$Ho_9Mg_{26}Zn_{65}$.

*Local coordination*

From the final refinements quite a number of different sorts of atoms and their coordination geometries result. There emerge 7 and 31 orbits in the "1/1" and "2/1" model, respectively. Most of them show FK type "normal"[37] coordinations (all P, Q, R and X: 85%). Although some of them are distorted in a way, that the strict geometric criteria for *tcp* structures[42] are not fulfilled. Not FK type coordinations have $\alpha^1$ (CN11; environment of the void cluster centers) and $\delta^Z$ (CN13; glue atom). 46% of all atoms are coordinated icosahedrally (CN12; FK polyhedron X). See Figure 7 for a view of the first coordination shells.

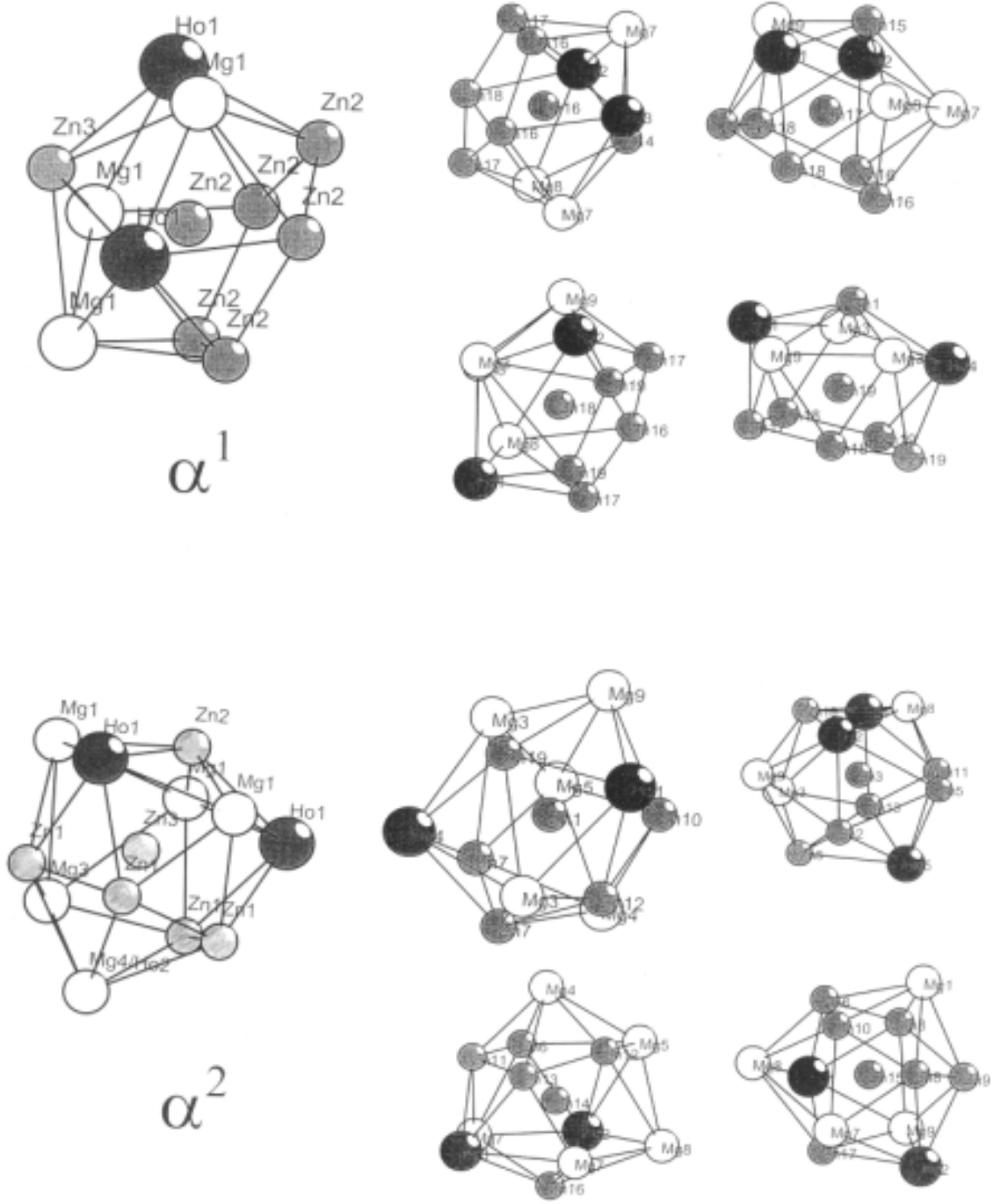

Figure 7 (continued)

$\alpha^3$

$\alpha^0$

Figure 7 (continued)

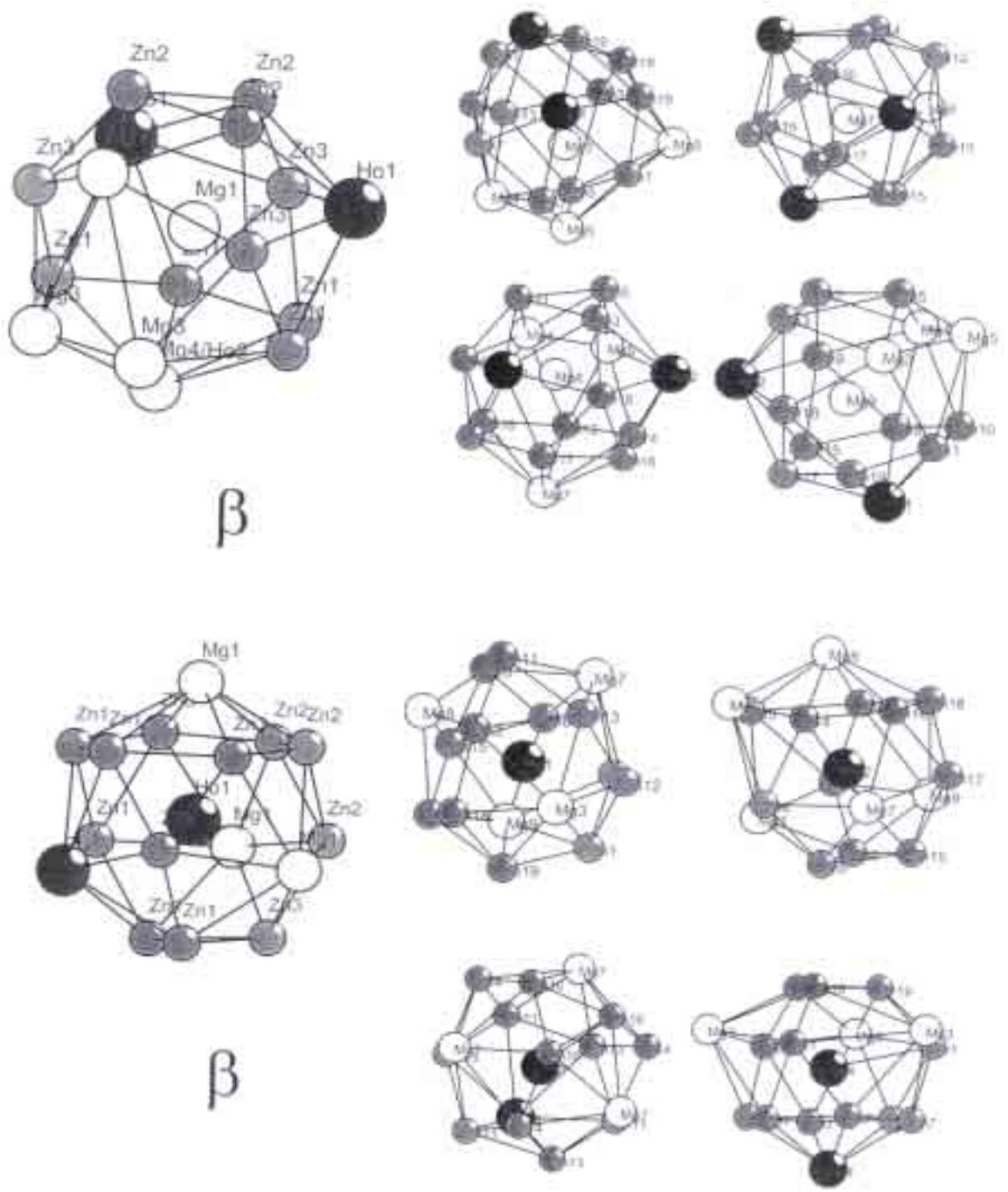

Figure 7 (continued)

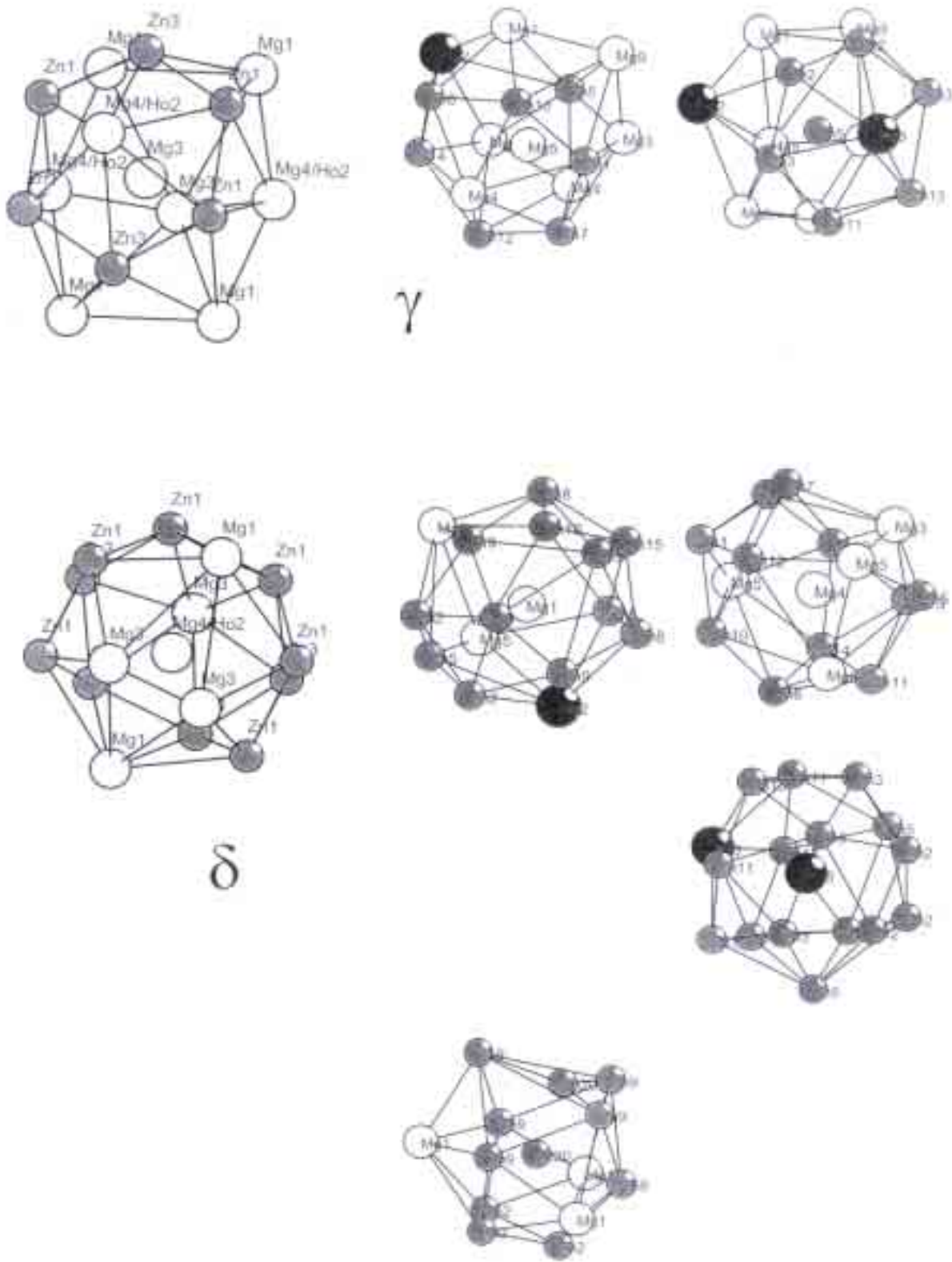

Figure 7. Local coordinations in *fci*-Ho$_9$Mg$_{26}$Zn$_{65}$. Left hand atoms from the "1/1" model, right hand (approx. half scale) the corresponding from the "2/1" model. Central atoms $\alpha^0$, $\alpha^1$, $\alpha^2$, $\beta$, $\gamma$, and $\delta$ denote atomic orbits corresponding to their structural function in the cluster. There is no corresponding glue atom $\delta^Z$ to Zn20 in the "1/1" model.

*Distortion of the structural elements*

There is a tendency towards greater distortion of the local coordination poyhedra (and therfore the clusters) switching from the "1/1" to "2/1" model. One could speculate whether this is continued for higher approximant models and thus the quasicrystal. Either the quasicrystal itself is built of highly symmetrical clusters and the above observation was an artifact of not appropriate symmetry constraints in the models. Or its local symmetries in fact are low, the symmetry of the clusters then is only near 2/*m*-3-5. Locally distorted but high long-range symmetries are commonly realized in crystalline giant cell structures, for a prominent example see Samsons β-$Al_3Mg_2$ (cF1168, *Fd-3m*, *a* = 28.24Å)[59].

*Average values*

In Figure 8 the average interatomic distances of each atom to its coordinating neighbours <*d*>, are plotted *vs.* the coordination number. Both in the "1/1" model as in the "2/1" model there is a trend of increasing <*d*> for larger CNs. That behaviour is common for crystalline intermetallic structures, compare the values for Laves´ $MgZn_2$[24]: <*d*>(Mg, CN16) = 3.098Å, <*d*>(Zn1, CN12) = 2.848Å and <*d*>(Zn2, CN12) = 2.841Å. The only exception is the "void" at the center of the clusters $α^0$. This again indicates that there is indeed *no* atom present in the center. The average distances in the Ho containing quasicrystal lie above the line given by $MgZn_2$, reflecting the slightly larger radius of Ho compared to Mg (see section IV).

Average coordination numbers were calculated for "2/1"-$Ho_{10.6}Mg_{24.7}Zn_{64.7}$. They are given in Table 6 and fit quite well the EXAFS data for *fci*-Mg-Zn-*RE* (*RE* = Dy, Y)[28].

| „average" atom | *N*(neighbours) Ho | *N*(neighbours) Mg | *N*(neighbours) Zn | $\Sigma N_i$ = <CN> |
|---|---|---|---|---|
| Ho | 0.33 | 3.67 | 12.00 | 16.00 |
| Mg | 1.57 | 3.14 | 10.71 | 15.42 |
| Zn | 1.96 | 4.09 | 5.85 | 11.90 |

Table 6. Number of neighbours and average coordination numbers in "2/1"-$Ho_{10.6}Mg_{24.7}Zn_{64.7}$.

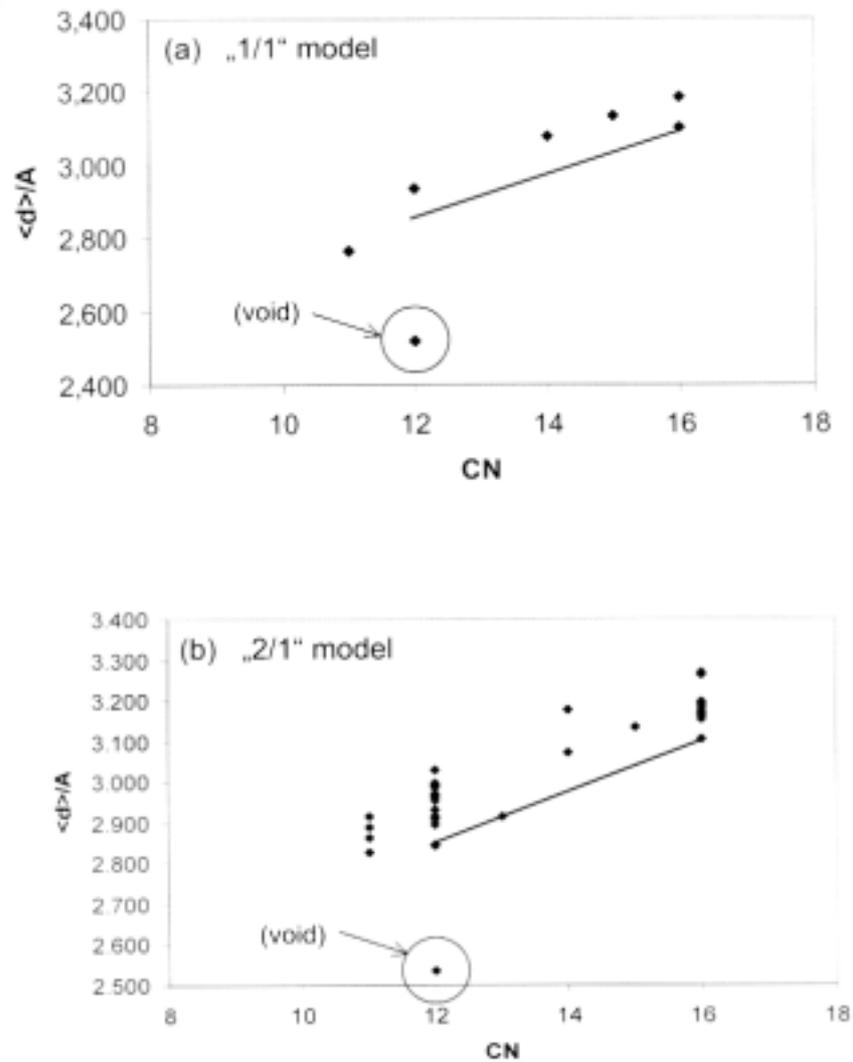

Figure 8.  Average coordinating distances $<d>$ are increasing with the coordination number CN; (a): "1/1", (b) "2/1" model; straight line: $MgZn_2$

*Structural function of the atoms*

As shown in Figure 7, the coordination polyhedra can be grouped according to their structural functions. The strong local resemblance of the "1/1" and "2/1" models can be seen clearly. Figure 9 shows the basic Bergman cluster in *fci*-$Ho_9Mg_{26}Zn_{65}$, as extracted from the "1/1" model: A central void ($\alpha^0$) is surrounded by 12 Zn-atoms ($\alpha^1$, inner icosahedron, $r = 2.5..3$Å), 8 Ho- and 12 Mg-atoms (β, pentagon dodecahedron/cube, $r = 4.5..5$Å), 12 Zn-atoms ($\alpha^2$, outer icosahedron, $r = 5..5.5$Å) and finally 48 Zn- and 12 Mg-atoms ($\alpha^3$ and γ, soccer ball, $r = 6.5..7.5$Å).

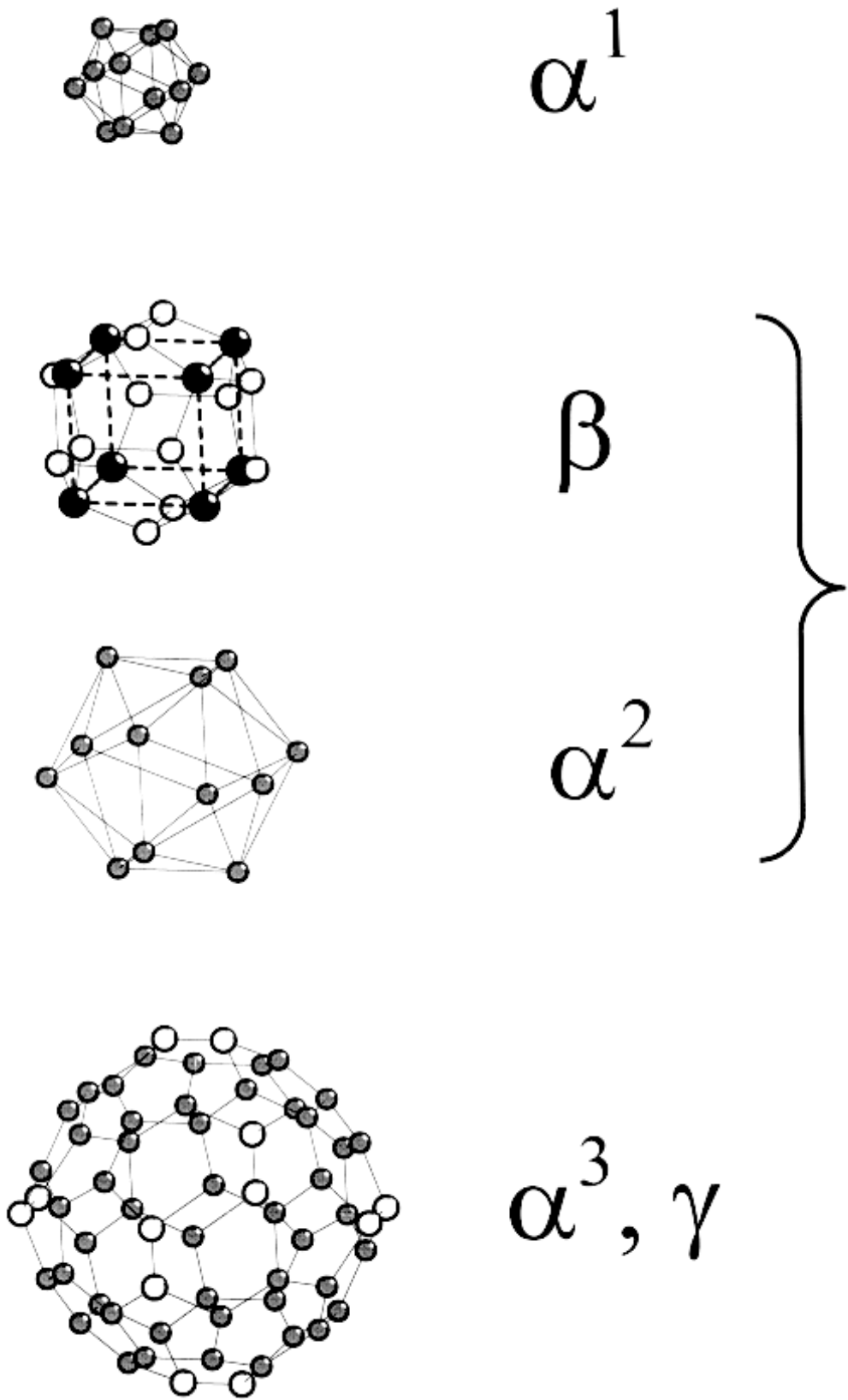

Figure 9.  Basic structural unit in *fci*-$Ho_9Mg_{26}Zn_{65}$ is the 104-atom Bergman cluster $\{[\text{void}]_1Ho_8Mg_{24}Zn_{72}\}$, here extracted from the "1/1" model. The 3 concentric shells [1. = $12\alpha^1$, 2. = $(20\beta+12\alpha^2)$, 3. = $(48\alpha^3+12\gamma)$] are drawn seperatedly for better recognition. It is $r(1.) = 2.5..3$Å, $r(2.) = 4.5..5.5$Å and $r(3.) = 6.5..7.5$Å.

In the "2/1" model there are 4 additional glue atoms per cluster (3×Zn20, $\delta^Z$ and 1×Ho5, $\delta^Y$) which are not present in the "1/1" model. Ho5 in the "2/1" model correponds exactly to the Ho part of the Mg4/Ho2 site in "1/1"-Ho$_{12.0}$Mg$_{28.0}$Zn$_{60.0}$. Also there are 3 more $\alpha^3$ atoms needed per cluster than in the "1/1" model. This is due to the different connecting scheme of the clusters: in "1/1" every cluster is connected to 8 neighbouring clusters sharing common hexagonal faces (*i.e.* 6 common atoms). In "2/1" there are only 7 next clusters connected that way — the missing face, because no longer shared, has to be refilled by 6/2 = 3 $\alpha^3$ atoms.

*"Odd" features*

The following features in our models would *not* be expected for ordinary intermetallic structures:

(i) short bonds (2.2 or 2.4Å) in the soccer ball: Mg3-Mg3 in the "1/1" model correspond to Mg5-Zn5 in "2/1". These all are atoms shared by two clusters. In the latter model, the glue atoms Mg4 and Zn20 suffer from one short contact per atom each (2.5 and 2.2 Å).

(ii) too large temperature factors are obseved in "2/1" for Zn3, Zn5 and Ho3 ($U_{eq}$ = 11.7, 12.4 and 15.2 × 10$^{-2}$Å$^2$, respectively). The Zn5 and Ho3 positions could be described as partially filled with Mg. This would be allowed by their coordination numbers (14 and 16). Anyway, we do not want to apply splitted occupation factors here intentionally. Because in the next higher approximant this *might* (and finally for the quasicrystal: *should*) resolve as shown above for the Ho glue atoms, ascending from "1/1" to "2/1". Interestingly, Zn3 in 2/1-Al-Mg-Zn is suffering from a strong anisotropic temperature factor corresponding to a large $U_{eq}$ of Zn3 in "2/1"-Ho$_{10.6}$Mg$_{24.7}$Zn$_{64.7}$.

Keeping in mind that we did refine our periodic *approximant* models using *quasicrystal* diffraction data, the local concordance of *fci* to "2/1" cubic is amazing. The only *intra* cluster irregularity affects the Ho cubes (β atoms) and might be due to the ordering of orientations of the cubes which is not perfect as discussed in section IV; see Figures 4 and 11. But most of the above *odd* features of the virtual approximant models are observed *inter* cluster. Thus we believe that they are a major key to higher virtual approximant models or, finally, to the long range quasicrystal structure.

*Cluster connecting scheme*

Figure 10 provides the connecting schemes of the clusters: The centers in the "1/1" model are arranged like the atoms in the *bcc*-W (cI2, *Im-3m*) structure. In terms of Henleys CCT[55], this is a (68) node: 6 *b*-bonds along 2fold icosahedral axes and 8 *c*-bonds along 3fold axes are meeting here. On the other hand, the arrangement of the clusters in the "2/1" model represents a (67) node: 6 *b*-bonds and 7 *c*-bonds are meeting in one point. (This is exactly the arrangement of the S atoms in pyrite, FeS$_2$, cP12, *Pa-3*).

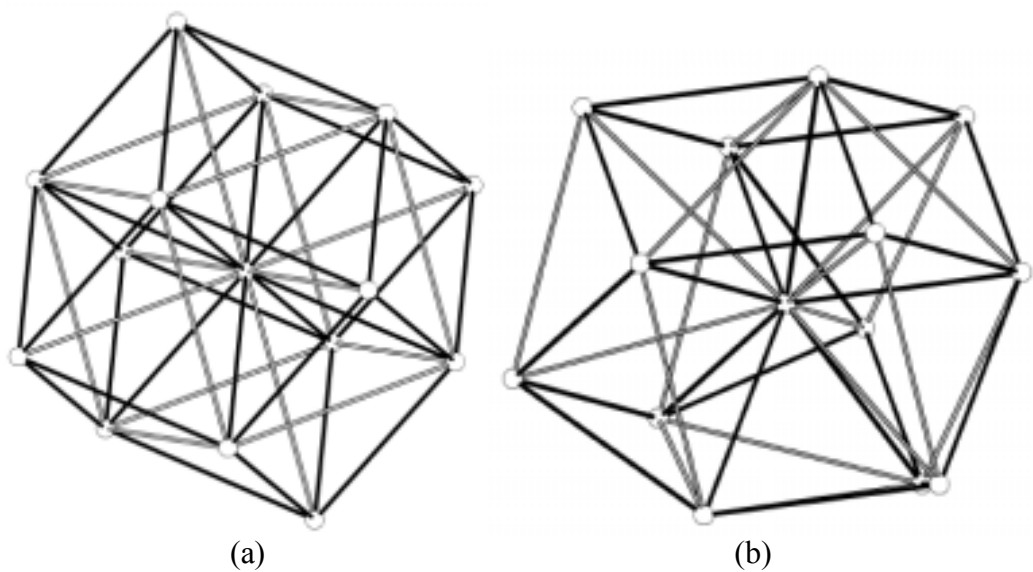

Figure 10.  Cluster centers arranged according to CCT schemes; *b*-bonds: open sticks, *c*-bonds: filled sticks; (a): (68) node environment in "1/1"-Ho-Mg-Zn, (b): (67) node environment in "2/1"-Ho-Mg-Zn

The space filling CCT comprises 4 different canonical cells, called A, B, C and D. Their edge lenghts are the above "bonds" between the cluster centers: $b$ (= $a_{1/1}$(3D) ≈ 14Å) and $c$ (= ½ √3 $b$ ≈ 12Å). The cubic unit cell for an 1/1 approximant consists of 24 A cells, the cubic unit cell of a 2/1 approximant of 24 A, 4 B and 4 C cells. They can be seen as fractions of the "coordination polyhedron" for the cluster packings in Figure 10. The 3/2 approximant then should comprise 72 A cells, 32 B, 32 C and 8 D cells, packed by another arrangement of (67) and (66) nodes. Unfortunately, there is no CCT for the quasicrystal found yet. For details see ref. 55. Nevertheless, our refinements strongly suggest that the *fci*-Ho-Mg-Zn quasicrystal is more like a (67) node than a (68) node arrangement of Bergman clusters. To get an impression, the latter is shown in soccer ball style in Figure 11.

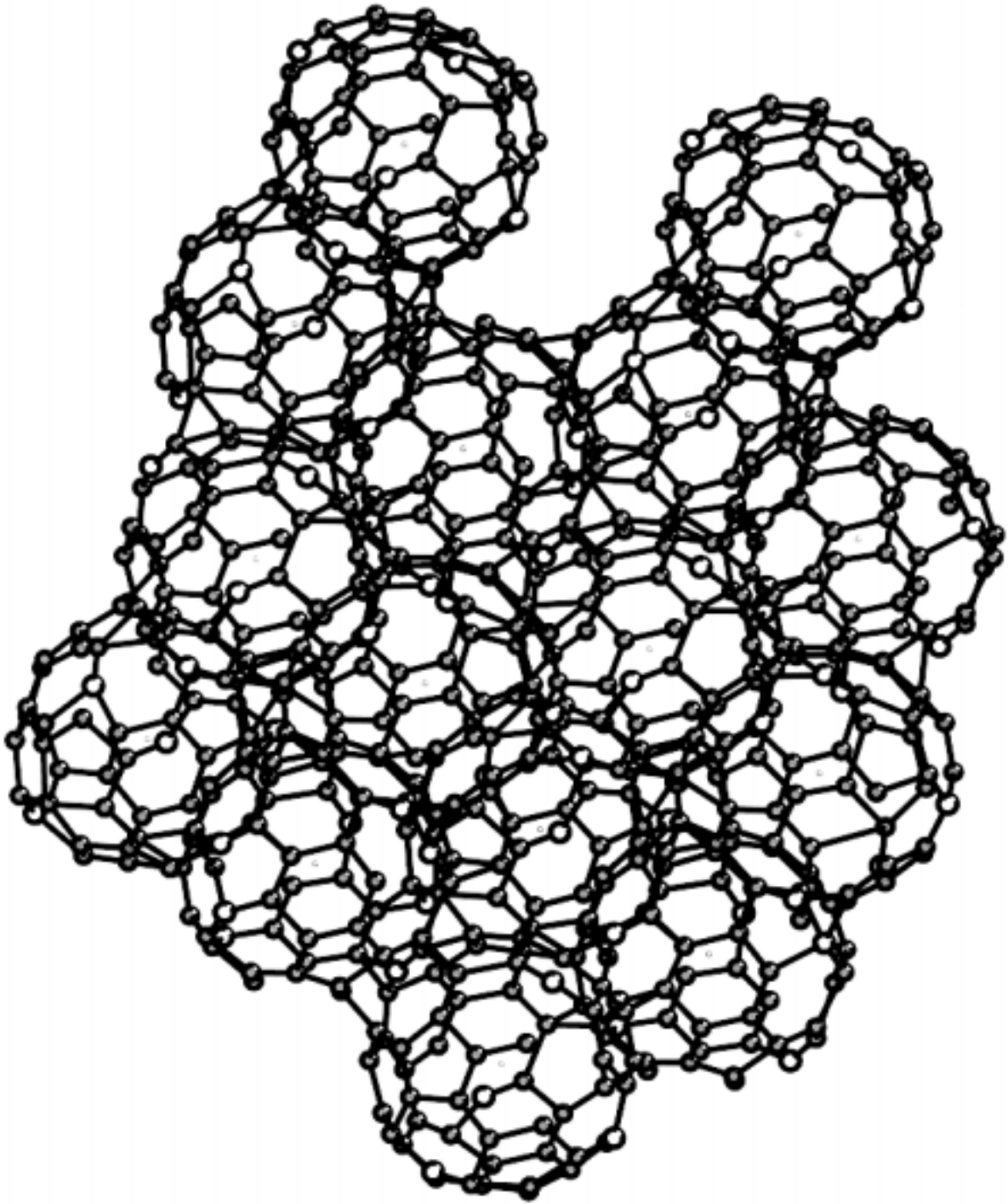

Figure 11. Cluster arrangement in the "2/1" model in soccer ball style. Only the $\alpha^3$ and $\gamma$ atoms are shown. The cores of the clusters cannot be seen, glue atoms are omitted in the drawing.

*Note on the term "cluster"*

Note that the term "cluster" for the structural units in intermetallic phases should not be mistaken with the well known "moleclular clusters" *e.g.* in solid $C_{60}$[60] or the $Mo_6S_8^{2-}$-cluster in Chevrel phases[61]. Here, atoms *intra* cluster are connected covalently. These isolated clusters then are *inter*connected by van der Waals or ionic interactions, leaving spatial gaps open and *e.g.* they can be isolated in an appropriate solvent. In our intermetallic *i* quasicrystal, the local coordinations of *all* atoms are almost isotropic. Spatial gaps *inter* cluster are filled by glue atoms. The only void in the icosahedral phase is the center of the cluster, not offering enough space for, say, the small Zn atom.

Yet, there seems to be superposition between local packing criteria (atoms regarded as hard spheres of slightly different size) and a (probably electronic) driving force to build icosahedral units of "molecular" dimensions (104 atom Bergman cluster: $\{[void]_1Ho_8Mg_{24}Zn_{72}\}$-unit; $e/a = 2.08$). Regarding the condensation of the clusters in the structure, a more realistic cluster formula (not accounting for glue atoms, though) would read $\{[void]_1Ho_8Mg_{(12+2b)}Zn_{(84-2b-3c)}\}$; b and c being the numbers of the connecting *b*- and *c*-bonds.

*Holmium partial structure*

See Figure 12 for a view of the Ho partial structure of "2/1"-$Ho_{10.6}Mg_{24.7}Zn_{64.7}$. It essentially consists of *intra* cluster $Ho_8$-cubes (Ho1 to Ho4; 89% of all Ho atoms). The average cube edge length is 5.4Å which fits well the shortest value derived from the 6D-refinement in ref. 31. The cubes are tilted with respect to each other minimizing the number of direct Ho-Ho contacts. One such contact remains inbetween neighbouring clusters in our "2/1" model: $d$(Ho4-Ho4) = 3.14Å. Another one is $d$(Ho3-Ho5) = 2.84Å. Ho5 acts as an *inter* cluster glue atom ($\delta^Y$). Ho5 itself is arranged like the cluster centers and forms another framework of (67) CCT nodes, shifted exactly (½ ½ ½), see also Figure 10b. Taking into account some misfit of observed and calculated PDFs at $r \approx 5.5$Å (Figure 6), we assume that in the proper quasicrystal a similar arrangement is realized avoiding Ho-Ho contacts < 5Å. On the quasicrystals long range scale, order-disorder phenomena as observed for *i*-$Mg_{30+x}Zn_{60}RE_{10-x}$, $x = 0..5$,[18] seem to be related to Mg substitution of the *RE* partial structure.

Basically, the location of the *RE*-atoms in *fci*-Mg-Tb-Zn in the 2nd and 5th shells as given in ref. 33 can be reproduced, though Ho in the shell (2a.) in our models is not arranged icosahedrally. This would correspond to the $\alpha^2$ site (shell 2b.) which has only CN12 and was too small to accommodate a large *RE* atom. The fifth shell should correspond to our glue atom position $\delta^Y$, 7.5..8Å distant from the cluster center. In fact, we do not have more than 3 shells in our Bergman cluster, the 3rd shell already is shared with neighbouring clusters. Nevertheless, in the "1/1" model one could interpret a 4th shell of Mg and Ho atoms forming a triacontahedron which can be related to an Ammann tiling.[62] The Ammann tiling consists of two sorts of "golden" rhombohedra, an oblate and a prolate one (termed OR and PR). Following certain matching rules, space can be filled with a 3D analogon of the 2D Penrose tiling. The tiling edge length is identical to the quasilattice constant, here $a(6D) = 5.18$Å. But the triacontahedron is widely interpenetrating with neighbouring clusters already (sharing one OR) and is not required to build the cluster decorated CCT model.

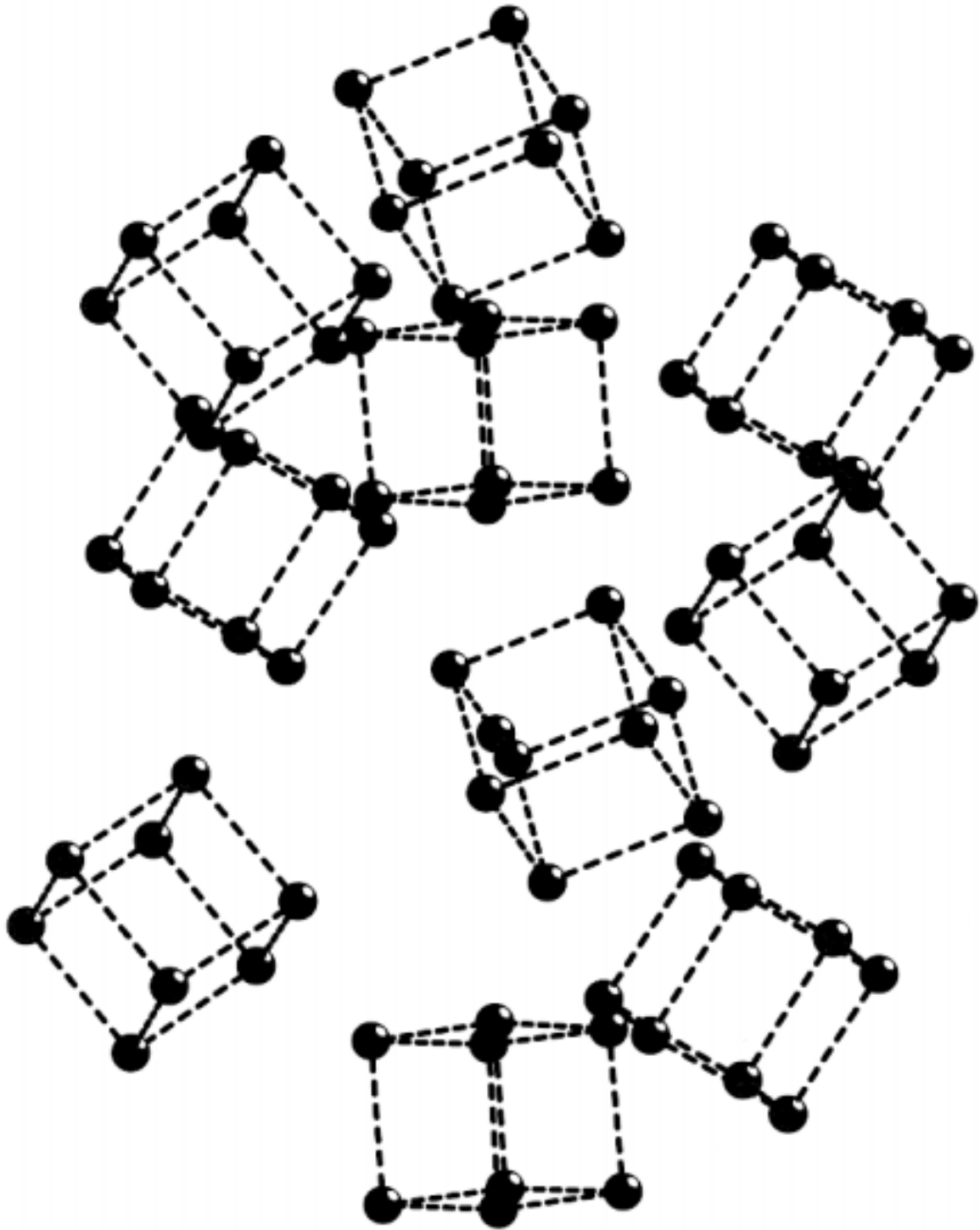

Figure12.  Holmium partial structure in the "2/1" model. In an arbitrary selection of clusters, only β-atoms (Ho1 to Ho4) forming the cubes of characteristic edge length 5.4Å are shown. Ho5 glue atoms (11% of all Ho atoms; $\delta^Y$) were omitted for clarity.

*Comparison to other (approximant) structures*

There is a very close relationsship between *i*-Ho-Mg-Zn and *i*-Mg-Zn-*B* (*B* = Al,Ga) and its approximants. As shown above, this fact was the key to the medium range structure solution of *fci*-Ho-Mg-Zn. Their local (almost) FK type structure is dominated by atom size effects. Starting from a hypothetical binary Mg-Zn phase, Zn(II) can be substituted by Al(III) or Ga(III); Mg(II) by *RE*(III). That way more than 2 valence electrons per atom (*e*/a) are introduced. In the Al or Ga case this can be done up to *e*/a = 2.5. The icosahedral phase has *e*/a ≈ 2.1, higher *e*/a values determine the formation of rational approximants[63]. In the *RE* case a maximum of 12 at% *RE* is known form the preparative investigations, yielding an *e*/a = 2.12. Thus corroborating ref. 63, no rational approximants have been observed here so far. From a structural point of view, higher *RE* contents would mean direct *RE*-*RE* contacts and it can be speculated that this was not compliant with the the space offered by the FK type polyhedron P (CN16). Note, in $HoZn_3$ (oP16, *Pnma*)[64], in *A*-$YZn_5$ (hP36, $P6_3/mmc$)[20] and in Z-$Mg_{28}Y_7Zn_{65}$ (**5**), Ho and Y are coordinated with CNs 17, 17 and 18, respectively. This again reflects the slightly larger radii of *RE* compared to Mg.

Our results in detail approve the Bergman type cluster as the only larger structural unit in *fci*-Ho-Mg-Zn, demarcating it from the Mackay type clusters as found *e.g.* in α-Al-Si-Mn[35]: There, the cluster center is surrounded by 12 Al-atoms, further 30 Al-atoms (icosidodecahedron) and 12 Mn-atoms (icosahedron): the latter 42 atoms form a so-called Mackay icosahedron (54 atoms in total).

Large clusters with *e.g.* 9 shells (*d* ≈ 20Å, 362 atoms) as in 2/1-Al-Mn-Pd-Si (cP513, *Pm*-3, *a* = 20.211Å)[65] are clearly not present in *fci*-Ho-Mg-Zn. The quite small size of the 3 shell Bergman cluster (*d* ≈ 15Å, 104 atoms) may be responsible for different electronic behaviour compared to Al-Mn-Pd[9]. Also, this fact might have given rise to speculations whether clusters *at all* are present in *fci*-Mg-Zn-*RE*[29].

Recently, a new icosahedral phase was discovered in the Mg-Sc-Zn system.[66] Its 1/1-approximant should be the compound $Sc_3Zn_{17}$ (cI160, *Im*-3, *a* = 13.852Å).[67] In contrast to *fci*-Ho-Mg-Zn the constituing cluster is built of a central void surrounded by 20 Zn-atoms (pentagon dodecahedron), 12 Sc-atoms (icosahedron), 30 Zn-atoms (icosidodecahedron) and 60 Zn-Atoms (soccer ball).

The structural similarities to the other ternary phases in the Mg-Zn-*RE* systems (**4**) to (**8**) are, if present, restricted to the local coordinations. For example Zn-atoms are frequently centers of distorted icosahedra. This was already recognized comparing the PDFs (see section IV); a detailed analysis may be subject of future discussions.

## VII. CONCLUSION

Quantitative analysis of the atomic pair distribution function (PDF) from X-ray powder diffraction data for the first time revealed the detailed local and medium range atomic structure of a quasicrystal. The strategy was (i) to use the PDF as a fingerprint for comparison to known crystalline structures. Once having found a good match, (ii) construction of "virtual" approximants in 3D real space follows. The local structure then can be (iii) refined using the PDF again. For other quasicrystals with real existing approximants, step (i) is redundant.

In *fci*-$Ho_9Mg_{26}Zn_{65}$, the local environments were determined as predominately FK-type (CNs 12, 14, 15 and 16). Basic structural unit is the 104-atom Bergman cluster comprising 3 shells. Ho atoms all have CN16, they are situated in the 2nd shell or act as glue atoms. The Bergman clusters decorate Henleys canonical cell tiling (CCT). A medium range arrangement of (67) CCT nodes seems to approximate the quasicrystalline long range order. In the "virtual" approximant models few too short bond lengths only occur where the clusters are interconnected. This indicates the limit of the models and might be a key for the construction of higher approximant models or the quasicrystal itself.

Therefore, the *fci*-Mg-Zn-*RE* quasicrystal structures may be considered as a compromise between local packing criteria (Frank-Kasper principle: atoms regarded as hard spheres of slightly different size) and an electronic driving force ($e/a \approx 2.1$; similarity to the concept of Hume-Rothery) to build icosahedral units of "molecular" dimensions. The latter then are packed quasiperiodically, all local coordinations are kept almost isotropic again by inserting some glue atoms.

In the future, an analogous treatment of *si*-Ho$_{10}$Mg$_{14}$Zn$_{76}$ (**2a**)[17] might give more insight in the long range order of the clusters. A virtual approximant "3/2"-Ho-Mg-Zn (cP2888, *Pa*-3, $a = 37$Å) could be constructed and refined using DISCUS[46]. The symmetry restrictions of *Pa*-3 should be replaced by "molecular" symmetry: 2/*m*-3-5 acting locally on the clusters in *P*1 should exclude (or confirm) local distortions. Better resolution by analysis of synchroton powder diffraction data (higher $Q_{max}$) may deepen the insight into the local structure and so indirectly provide information on the long range order.

The similarities of the structural features in perodic 2/1-Al-Mg-Zn and quasiperiodic *fci*-Ho-Mg-Zn are a strong argument for the *n*D cut-and-projection method. They may give rise for improved constructions (and refinements) in 6D space, see the recent lifting of 1/1-(Al,Si)-Cu-Fe to 6D *i*-(Al,Si)-Cu-Fe.[68] Refinements coupling both 1D PDF and 6D approaches may elaborate quasicrystal structure analysis.

Finally, the construction of an icosahedrally quasiperiodic CCT and its decoration according to our findings should provide an outright description of *fci*-Mg-Zn-*RE* in 3D.

## ACKNOWLEDGEMENTS

Financial support by the DFG (Schwerpunktprogramm 1031 "Quasikristalle") is gratefully acknowledged. We like to thank Walter Steurer for his kind suggestion to go along with the PDF refinement technique, Thomas Proffen for his "on-line" assistance with the DIFFUSE software via the WWW and Klaus-Dieter Luther for his technical skills to upgrade the Guinier diffractometer. One of us (S.B.) thanks Birke Brühne for her support to re-enter quasicrystal research again.

________________________________